\documentclass[conference,compsoc]{IEEEtran}
\usepackage[nocompress]{cite}
\usepackage{acronym}
\usepackage{needspace}
\acrodef{KV}{key-value}
\acrodef{MAE}{mean absolute error}
\acrodef{MIG}{multi-instance GPU}
\acrodef{MoE}{mixture-of-experts}
\acrodef{NVML}{NVIDIA Management Library}
\acrodef{MAC}{multiply-accumulate}
\acrodef{LUT}{lookup table}
\usepackage{amsmath,amssymb}
\usepackage{array,tabularx,booktabs}
\usepackage[caption=false,font=footnotesize]{subfig}
\usepackage{xurl}
\usepackage[scaled=0.92]{helvet}
\usepackage{fontawesome5}
\usepackage{pgfplots}
\usepgfplotslibrary{groupplots,fillbetween}
\usetikzlibrary{positioning,arrows.meta,calc,fit,backgrounds,shadows.blur}
\definecolor{tolblue}{HTML}{4477AA}
\definecolor{tolred}{HTML}{EE6677}
\definecolor{tolgreen}{HTML}{228833}
\definecolor{tolyellow}{HTML}{CCBB44}
\definecolor{tolcyan}{HTML}{66CCEE}
\definecolor{tolpurple}{HTML}{AA3377}
\definecolor{tolgrey}{HTML}{BBBBBB}
\colorlet{ink}{black!78}
\colorlet{secretc}{tolred}
\colorlet{obsc}{tolblue}
\colorlet{physc}{black!55}
\colorlet{posc}{tolgreen}
\colorlet{negc}{black!40}

\tikzset{
  every picture/.append style={font=\sffamily\footnotesize, line width=0.6pt, draw=ink, text=ink},
  card/.style={draw=#1!85!black, fill=#1!9, rounded corners=3pt, line width=0.7pt,
               align=center, inner sep=4pt, minimum height=10mm},
  shadowed/.style={blur shadow={shadow blur steps=6, shadow xshift=0.5pt, shadow yshift=-0.8pt,
                   shadow blur radius=1.6pt, shadow opacity=22}},
  flow/.style={-{Stealth[round, length=2.2mm, width=1.8mm]}, ink, line width=0.7pt},
  flowneg/.style={flow, negc, dash pattern=on 2pt off 1.5pt},
  band/.style={fill=black!3.5, draw=none, rounded corners=4pt},
  badge/.style={circle, inner sep=0.6pt, font=\scriptsize, text=white, fill=#1},
  icon/.style={font=\large, text=#1},
  statnode/.style={anchor=north west, align=left, font=\sffamily\scriptsize, fill=white, fill opacity=0.85, text opacity=1, inner sep=1.5pt},
}

\pgfplotsset{
  compat=1.18,
  paperplot/.style={
    axis lines*=left, axis line style={ink, line width=0.5pt},
    tick style={ink, line width=0.4pt}, tick align=outside, major tick length=2pt,
    ymajorgrids, grid style={black!10, line width=0.3pt},
    tick label style={font=\sffamily\scriptsize}, /pgf/number format/assume math mode=true, label style={font=\sffamily\footnotesize},
    legend style={font=\sffamily\scriptsize, draw=none, fill=none},
    every axis plot/.append style={line width=0.9pt},
  },
}

\newcommand{\ResultAgeAuthors}{60}
\newcommand{\ResultAgeBaseline}{11.333}

\newcommand{\ResultAgeCalibrationMax}{.903}
\newcommand{\ResultAgeCalibrationMin}{.867}
\newcommand{\ResultAgeClean}{7.559}

\newcommand{\ResultAgeCoordinatesMax}{.889}
\newcommand{\ResultAgeCoordinatesMin}{.688}
\newcommand{\ResultAgeCorrelation}{.895}
\newcommand{\ResultAgeCuda}{12.6}
\newcommand{\ResultAgeDate}{2026-10-23}
\newcommand{\ResultAgeDriver}{560.35.03}
\newcommand{\ResultAgeGain}{3.56}
\newcommand{\ResultAgeGainCiHigh}{5.03}
\newcommand{\ResultAgeGainCiLow}{2.04}
\newcommand{\ResultAgeInvalidSamples}{3}
\newcommand{\ResultAgeP}{\ensuremath{4.000\times10^{-5}}}
\newcommand{\ResultAgePairs}{336}
\newcommand{\ResultAgePhysical}{7.778}
\newcommand{\ResultAgePhysicalCiHigh}{9.78}
\newcommand{\ResultAgePhysicalCiLow}{5.96}

\newcommand{\ResultAgePowerLimit}{300}

\newcommand{\ResultAgeRecords}{672}
\newcommand{\ResultAgeSamples}{1,209,600}
\newcommand{\ResultAgeSecondsPerAuthor}{320}
\newcommand{\ResultAgeSham}{11.346}

\newcommand{\ResultAgeTorch}{2.8.0+cu126}
\newcommand{\ResultAgeTotalHours}{7.47}
\newcommand{\ResultAgeTransitionDuration}{.971}

\newcommand{\ResultAgeWorkTransition}{.845}
\newcommand{\ResultCostFixedBridgeCaptures}{20,480}
\newcommand{\ResultCostFixedBridgeSeconds}{65.7}
\newcommand{\ResultCostFixedProfileCaptures}{327,680}
\newcommand{\ResultCostFixedProfileSeconds}{1,054.7}
\newcommand{\ResultCostFixedVictimCaptures}{409,600}
\newcommand{\ResultCostFixedVictimSeconds}{1,321.8}
\newcommand{\ResultCostNaturalBridgeCaptures}{81,920}
\newcommand{\ResultCostNaturalBridgeSeconds}{267.0}
\newcommand{\ResultCostNaturalProfileCaptures}{1,310,720}
\newcommand{\ResultCostNaturalProfileSeconds}{4,259.6}
\newcommand{\ResultCostNaturalVictimCaptures}{1,638,400}
\newcommand{\ResultCostNaturalVictimSeconds}{5,321.1}
\newcommand{\ResultCwBridgePeople}{32}
\newcommand{\ResultCwConditionalP}{1/200001}
\newcommand{\ResultCwDonor}{283}
\newcommand{\ResultCwFixedConditionBlockRange}{50–51/64}
\newcommand{\ResultCwFixedConditionP}{\ensuremath{1.25\times10^{-7}}}

\newcommand{\ResultCwFixedDirectConditionCount}{49/64}

\newcommand{\ResultCwFixedDirectJointCount}{33/64}

\newcommand{\ResultCwFixedDirectVaultCount}{42/64}

\newcommand{\ResultCwFixedJointP}{\ensuremath{6.54\times10^{-8}}}

\newcommand{\ResultCwFixedPairedConditionCount}{51/64}

\newcommand{\ResultCwFixedPairedJointCount}{34/64}

\newcommand{\ResultCwFixedPairedVaultCount}{42/64}

\newcommand{\ResultCwFixedShamConditionCount}{18/64}

\newcommand{\ResultCwFixedShamJointCount}{5/64}

\newcommand{\ResultCwFixedShamVaultCount}{15/64}

\newcommand{\ResultCwFixedVaultBlockRange}{40–42/64}
\newcommand{\ResultCwFixedVaultP}{\ensuremath{3.71\times10^{-6}}}
\newcommand{\ResultCwNaturalConditionBlockRange}{50–52/64}
\newcommand{\ResultCwNaturalConditionP}{\ensuremath{5.63\times10^{-11}}}

\newcommand{\ResultCwNaturalDirectConditionCount}{47/64}

\newcommand{\ResultCwNaturalDirectJointCount}{36/64}

\newcommand{\ResultCwNaturalDirectVaultCount}{48/64}

\newcommand{\ResultCwNaturalJointP}{\ensuremath{2.76\times10^{-10}}}

\newcommand{\ResultCwNaturalPairedConditionCount}{51/64}

\newcommand{\ResultCwNaturalPairedJointCount}{40/64}

\newcommand{\ResultCwNaturalPairedVaultCount}{49/64}

\newcommand{\ResultCwNaturalShamConditionCount}{11/64}

\newcommand{\ResultCwNaturalShamJointCount}{5/64}

\newcommand{\ResultCwNaturalShamVaultCount}{25/64}

\newcommand{\ResultCwNaturalVaultBlockRange}{48–49/64}
\newcommand{\ResultCwNaturalVaultP}{\ensuremath{9.65\times10^{-6}}}
\newcommand{\ResultGraniteConditionLayers}{14}
\newcommand{\ResultGraniteFraction}{0.149}
\newcommand{\ResultGraniteInterventionCombined}{\ensuremath{+9.56}}
\newcommand{\ResultGraniteInterventionCombinedHigh}{\ensuremath{+10.25}}
\newcommand{\ResultGraniteInterventionCombinedLow}{\ensuremath{+8.90}}
\newcommand{\ResultGraniteInterventionState}{\ensuremath{+7.47}}
\newcommand{\ResultGraniteInterventionStateHigh}{\ensuremath{+8.34}}
\newcommand{\ResultGraniteInterventionStateLow}{\ensuremath{+6.69}}
\newcommand{\ResultGraniteLateOne}{85.9}
\newcommand{\ResultGraniteLateTwo}{89.1}
\newcommand{\ResultGraniteOneCompliance}{254/256}
\newcommand{\ResultGraniteOneCondition}{68.8}
\newcommand{\ResultGraniteOneLateSelected}{24/32}
\newcommand{\ResultGraniteOneRule}{100.0}
\newcommand{\ResultGraniteOneSelected}{75.0}
\newcommand{\ResultGraniteStateElements}{5,505,024}
\newcommand{\ResultGraniteTwoCompliance}{252/256}
\newcommand{\ResultGraniteTwoCondition}{82.8}
\newcommand{\ResultGraniteTwoLateSelected}{21/32}

\newcommand{\ResultInterventionCombinedHigh}{\ensuremath{+1.911}}
\newcommand{\ResultInterventionCombinedLow}{\ensuremath{+1.716}}
\newcommand{\ResultInterventionCombinedMean}{\ensuremath{+1.811}}
\newcommand{\ResultInterventionCombinedPairs}{384/384}
\newcommand{\ResultInterventionFinal}{\ensuremath{+1.587}}
\newcommand{\ResultInterventionInitial}{\ensuremath{+2.275}}
\newcommand{\ResultInterventionLateHigh}{\ensuremath{+1.844}}
\newcommand{\ResultInterventionLateLow}{\ensuremath{+1.674}}
\newcommand{\ResultInterventionLateMean}{\ensuremath{+1.756}}
\newcommand{\ResultInterventionLatePairs}{384/384}
\newcommand{\ResultInterventionPBound}{10^{-6}}
\newcommand{\ResultInterventionPositivePairs}{384}
\newcommand{\ResultInterventionPositiveProfiles}{32}
\newcommand{\ResultInterventionRecurrentHigh}{\ensuremath{+1.915}}
\newcommand{\ResultInterventionRecurrentLow}{\ensuremath{+1.715}}
\newcommand{\ResultInterventionRecurrentMean}{\ensuremath{+1.811}}
\newcommand{\ResultInterventionRecurrentPairs}{384/384}
\newcommand{\ResultInterventionShamHigh}{0.000}
\newcommand{\ResultInterventionShamLow}{0.000}
\newcommand{\ResultInterventionShamMean}{0.000}
\newcommand{\ResultInterventionShamPairs}{0/384}
\newcommand{\ResultLateStatusBehavior}{123/128}
\newcommand{\ResultLateStatusCount}{47/128}
\newcommand{\ResultLateStatusP}{.00647}
\newcommand{\ResultLateVerifyBehavior}{124/128}
\newcommand{\ResultLateVerifyCount}{45/128}
\newcommand{\ResultLateVerifyP}{.0133}
\newcommand{\ResultMechanismDuration}{.8571}
\newcommand{\ResultMechanismDurationP}{\ensuremath{9.999\times10^{-5}}}
\newcommand{\ResultMechanismMae}{.3091}
\newcommand{\ResultMechanismSham}{.0361}
\newcommand{\ResultMechanismShamP}{.8557}
\newcommand{\ResultMechanismShift}{1.0869}
\newcommand{\ResultMechanismTracking}{.9615}
\newcommand{\ResultMechanismTransition}{.6806}
\newcommand{\ResultMechanismTransitionP}{\ensuremath{9.999\times10^{-5}}}

\newcommand{\ResultNaturalAgeAuthors}{60}
\newcommand{\ResultNaturalAgeBaseline}{11.33}
\newcommand{\ResultNaturalAgeBatchLanes}{16}
\newcommand{\ResultNaturalAgeClean}{8.32}
\newcommand{\ResultNaturalAgeHistoryTokens}{256}
\newcommand{\ResultNaturalAgeInvalidSamples}{5}
\newcommand{\ResultNaturalAgeP}{.0245}
\newcommand{\ResultNaturalAgePairs}{336}
\newcommand{\ResultNaturalAgePhysical}{9.64}
\newcommand{\ResultNaturalAgePowerDurationRho}{.964}
\newcommand{\ResultNaturalAgeProbeRepeats}{20}
\newcommand{\ResultNaturalAgePublicTokens}{8}
\newcommand{\ResultNaturalAgeRecords}{672}

\newcommand{\ResultNaturalAgeSamples}{1,008,000}
\newcommand{\ResultNaturalAgeSamplesPerRecord}{1500}
\newcommand{\ResultNaturalAgeSessions}{88}
\newcommand{\ResultNaturalAgeSham}{11.02}
\newcommand{\ResultNaturalAgeSlotSeconds}{40}
\newcommand{\ResultNaturalBatchAnchorRhoMax}{.972}
\newcommand{\ResultNaturalBatchAnchorRhoMin}{.816}
\newcommand{\ResultNaturalBatchWorkDurationRho}{.982}
\newcommand{\ResultNaturalBatchWorkPowerRho}{.945}
\newcommand{\ResultNoiseOneConditionMax}{55.2}
\newcommand{\ResultNoiseOneConditionMin}{49.4}
\newcommand{\ResultNoiseOneRuleMax}{90.9}
\newcommand{\ResultNoiseOneRuleMin}{88.6}
\newcommand{\ResultNoiseTwoConditionMax}{41.7}
\newcommand{\ResultNoiseTwoConditionMin}{36.9}
\newcommand{\ResultNoiseTwoRuleMax}{71.1}
\newcommand{\ResultNoiseTwoRuleMin}{68.8}
\newcommand{\ResultOperatorError}{\ensuremath{1.77\times10^{-5}}}

\newcommand{\ResultQuantizationCorrelation}{.99426}
\newcommand{\ResultQuantizationNrmse}{.10630}
\newcommand{\ResultQwenFraction}{0.174}
\newcommand{\ResultQwenOneAllThreeFuture}{7.8}
\newcommand{\ResultQwenOneBothPrivateFacts}{36.7}
\newcommand{\ResultQwenOneCondition}{83.6}
\newcommand{\ResultQwenOnePing}{50.0}
\newcommand{\ResultQwenOneStatus}{43.0}
\newcommand{\ResultQwenOneVault}{43.0}
\newcommand{\ResultQwenOneVerify}{35.2}
\newcommand{\ResultQwenStateElements}{4,718,592}
\newcommand{\ResultQwenTwoAllThreeFuture}{10.2}
\newcommand{\ResultQwenTwoBothPrivateFacts}{29.7}
\newcommand{\ResultQwenTwoCondition}{89.1}
\newcommand{\ResultQwenTwoPing}{47.7}
\newcommand{\ResultQwenTwoStatus}{43.8}
\newcommand{\ResultQwenTwoVault}{33.6}
\newcommand{\ResultQwenTwoVerify}{40.6}
\newcommand{\ResultRetentionConditionControl}{21.1}
\newcommand{\ResultRetentionConditionFull}{42.2}
\newcommand{\ResultRetentionConditionP}{\ensuremath{3.47\times10^{-6}}}
\newcommand{\ResultRetentionPolicyControl}{25.8}
\newcommand{\ResultRetentionPolicyFull}{45.3}
\newcommand{\ResultRetentionPolicyP}{\ensuremath{4.17\times10^{-7}}}
\newcommand{\ResultRetentionTokensMax}{948}
\newcommand{\ResultRetentionTokensMin}{946}
\newcommand{\ResultRoutingMax}{465}
\newcommand{\ResultRoutingMin}{209}
\newcommand{\ResultRoutingResidual}{.90623}
\newcommand{\ResultRoutingResidualP}{\ensuremath{9.999\times10^{-5}}}
\newcommand{\ResultRoutingSingleton}{.99258}
\newcommand{\ResultSynthAgeBaseline}{11.30}
\newcommand{\ResultSynthAgeGainLoss}{216/84}
\newcommand{\ResultSynthAgeMae}{6.56}
\newcommand{\ResultSynthAgeP}{\ensuremath{6.60\times10^{-14}}}
\newcommand{\ResultSynthAgeWithinFive}{47.7}
\newcommand{\ResultSynthAuthors}{300}
\newcommand{\ResultSynthBaselineWithinFive}{24.0}
\newcommand{\ResultSynthComments}{7,823}

\newcommand{\ResultSynthesisLuts}{753}
\newcommand{\ResultSynthesisRegisters}{1,405}

\providecommand{\tightlist}{\setlength{\itemsep}{0pt}\setlength{\parskip}{0pt}}

\usepackage[hidelinks]{hyperref}
\hypersetup{pdftitle={Privacy Leakage from Recurrent State in LLM Inference},pdfauthor={Weimin Fu, Shikun Liu, Qiying Li, Yuntao Liu, Pan Li, Xiaolong Guo}}

\begin{document}
\raggedbottom

\title{Privacy Leakage from Recurrent State in LLM Inference}
\author{\IEEEauthorblockN{Weimin Fu\IEEEauthorrefmark{1}, Shikun Liu\IEEEauthorrefmark{2}, Qiying Li\IEEEauthorrefmark{1}, Yuntao Liu\IEEEauthorrefmark{1}, Pan Li\IEEEauthorrefmark{2}, Xiaolong Guo\IEEEauthorrefmark{1}}
\IEEEauthorblockA{\IEEEauthorrefmark{1}Lehigh University \quad \IEEEauthorrefmark{2}Georgia Institute of Technology\\ \textit{Preprint}}}
\maketitle

\begin{abstract}
Personalized LLM services place private facts and application instructions into a prompt, then retain representations of that context for later inference. Latent-state LLMs compress part of this history into fixed-size recurrent state. The information preserved by its learned updates is implicit, and later computation can expose it even when an observer receives neither the private text nor the responses. We investigate privacy leakage from recurrent state through partial-state probing and physical observation of its use. Our methodology separates recoverable information, persistence through public dialogue, causal influence on later decisions, and leakage through power. Experiments with released hybrid implementations show that a small state observation supports multiple predictions about private facts and response rules, and that personal attributes generalize to unseen authors. The information remains inferable after public dialogue and changes later decision margins. Complete-read power recovers private facts under a fixed execution schedule. On a partitioned GPU, an unprivileged account infers age from whole-device power during public requests; a separate state-isolation experiment establishes the recurrent contribution to this channel. These findings extend context confidentiality to the lifetime and execution of retained state. They identify access-control responsibilities for serving systems, telemetry restrictions for GPU operators, and data-dependent power leakage for accelerator designers.
\end{abstract}

\IEEEpeerreviewmaketitle

\section{Introduction}\label{sec:1}

Modern assistants personalize their answers by retaining information about each user across sessions. OpenAI's ChatGPT memory, introduced in 2024 and extended in 2025 to draw on a user's earlier chats, stores personal details and stated preferences so that later replies reflect them \cite{r32}. The personal context these services retain is sensitive, and it has been exposed in practice: in 2026 a single misconfigured backend made hundreds of millions of assistant messages from tens of millions of users publicly readable \cite{r33}. A deployed service also places this context, together with the private instructions that govern how it answers, into the prompt of each request. What a service holds about a user includes facts the user supplied, the rules the application enforces, and attributes the user never stated directly.

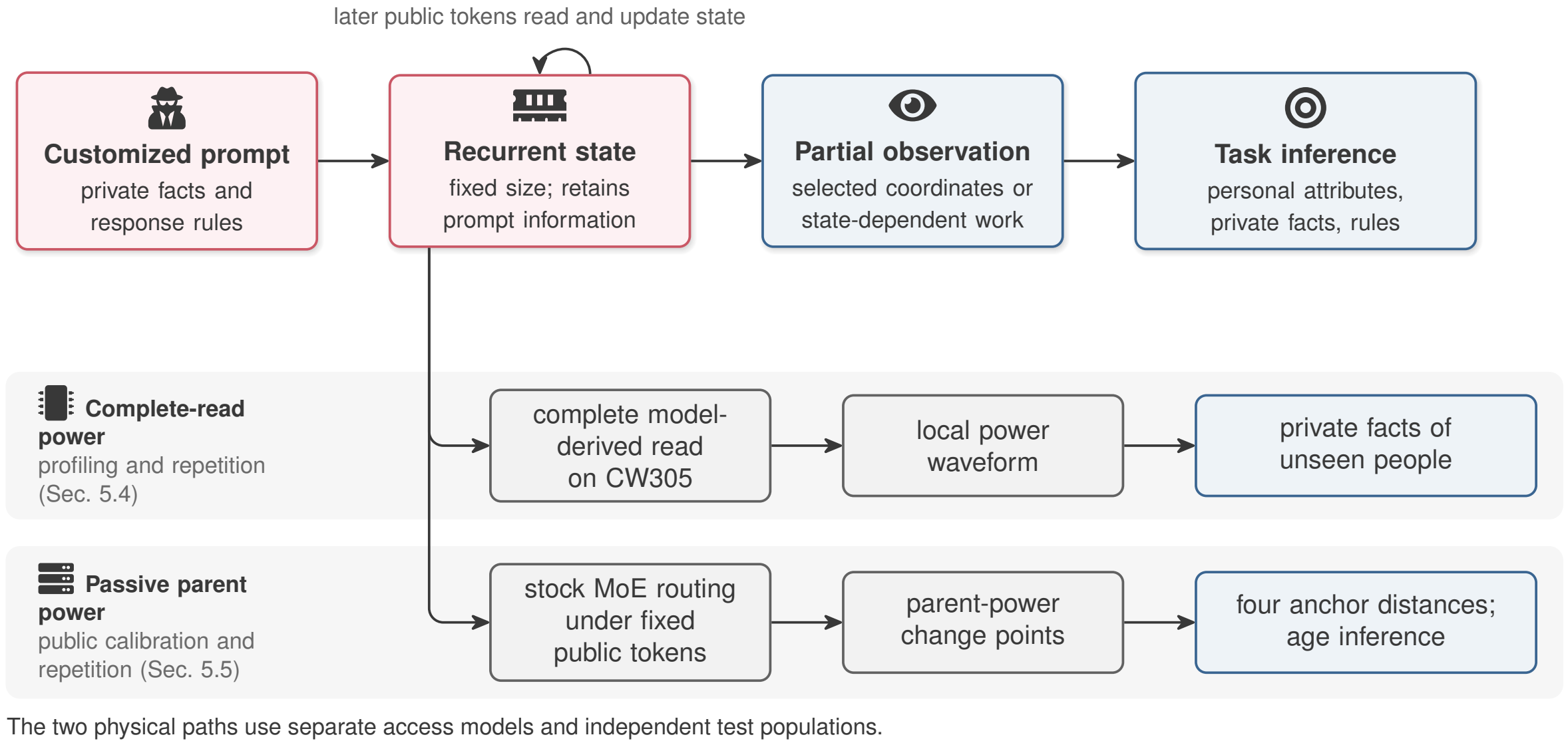
\begin{figure*}[t]
\centering
\begin{tikzpicture}[
  node distance=4mm and 7mm,
  step/.style={card=#1, text width=27mm},
  res/.style={card=#1, text width=31mm},
  lanelabel/.style={anchor=west, align=left, font=\sffamily\scriptsize, text width=25mm},
  sub/.style={font=\sffamily\scriptsize, text=ink!80},
]
\node[step=secretc, shadowed] (H) {{\large\faUserSecret}\\[2pt]\textbf{Customized prompt}\\[1pt]{\scriptsize private facts and response rules}};
\node[step=secretc, shadowed, right=of H] (S) {{\large\faMemory}\\[2pt]\textbf{Recurrent state}\\[1pt]{\scriptsize fixed size; retains prompt information}};
\node[step=obsc, shadowed, right=of S] (Z) {{\large\faEye}\\[2pt]\textbf{Partial observation}\\[1pt]{\scriptsize selected coordinates or state-dependent work}};
\node[res=obsc, shadowed, right=of Z] (T) {{\large\faBullseye}\\[2pt]\textbf{Task inference}\\[1pt]{\scriptsize personal attributes, private facts, rules}};
\draw[flow] (H) -- (S);
\draw[flow] (S) -- (Z);
\draw[flow] (Z) -- (T);
\draw[flow] ([xshift=5mm]S.north) arc[start angle=0, end angle=180, radius=2.5mm];
\node[sub, above=3.2mm of S.north, anchor=south] {later public tokens read and update state};

\node[card=physc, text width=25mm, below=14mm of S, xshift=9mm] (a1) {complete model-derived read on CW305};
\node[card=physc, text width=25mm, right=of a1] (a2) {local power waveform};
\node[res=obsc, right=of a2] (a3) {private facts of unseen people};
\node[card=physc, text width=25mm, below=6mm of a1] (b1) {stock MoE routing under fixed public tokens};
\node[card=physc, text width=25mm, right=of b1] (b2) {parent-power change points};
\node[res=obsc, right=of b2] (b3) {four anchor distances; age inference};
\foreach \r in {a,b} {
  \draw[flow] (\r1) -- (\r2);
  \draw[flow] (\r2) -- (\r3);
}
\begin{scope}[on background layer]
  \foreach \r/\t/\s/\i in {a/{Complete-read power}/{profiling and repetition (Sec.~5.4)}/\faMicrochip,
                           b/{Passive parent power}/{public calibration and repetition (Sec.~5.5)}/\faServer} {
    \node[band, fit=(\r1)(\r3), inner xsep=2.5mm, inner ysep=1.6mm] (band\r) {};
    \fill[black!3.5, rounded corners=4pt] ([xshift=-1mm]H.west |- band\r.north) rectangle (band\r.south east);
    \node[lanelabel] at ([xshift=1mm]H.west |- \r1) {{\normalsize\i}\enspace\textbf{\t}\\{\color{ink!75}\s}};
  }
\end{scope}
\foreach \n in {a1,b1} \draw[flow, rounded corners=4pt] ([xshift=-11mm]S.south) |- (\n.west);
\node[anchor=north west, font=\sffamily\scriptsize, inner sep=0pt, text=ink] at ([yshift=-3.5mm, xshift=-1mm]H.west |- b1.south)
  {The two physical paths use separate access models and independent test populations.};
\end{tikzpicture}
\caption{Confidential prompt information retained in the recurrent state of a latent-state LLM. Software experiments test partial-state inference, persistence, and later decisions. Complete-read power recovers a protected secret. On GPU, a state-isolated run attributes the routing signal to the recurrent state, and a natural-session service lets one unprivileged account infer a personal attribute from parent power. Each result uses its stated access conditions and held-out evaluation.}
\label{fig:1}
\end{figure*}

Recent hybrid LLM architectures provide another way to retain conversation history. Full-attention inference stores a key and value vector for every past token in a \ac{KV} cache whose size grows with context length \cite{r2}. To bound this cost, hybrid architectures replace the per-token storage in part of their layers with a fixed-size recurrent state; we call these models latent-state LLMs, the linear-attention and state-space hybrids among recent models \cite{r1,r2,r3,r34,r35}. Each token updates the state, mixing new information with retained history through learned gates and update rules. As long as the service carries this state forward, earlier input can influence later requests without being supplied again as text. An explicit text summary makes the retained wording available for inspection. Recurrent retention happens within numerical updates, without producing such a record of which private facts or instructions remain recoverable. A \ac{KV} entry identifies a token position; a recurrent-state entry combines contributions from many tokens. Removing a fact from the visible prompt does not itself remove its earlier contribution to a state that is still reused. The privacy question is what information remains recoverable from that state and from the later computations that read it.

Prior privacy attacks establish that protecting the input text alone does not protect every representation derived from it. Prompt-stealing attacks reconstruct prompts from shared-cache behavior \cite{r22} or a provider's plaintext \ac{KV} cache \cite{r23}. Embedding and activation attacks recover text from exchanged representations \cite{r6,r7,r21}. GPU side channels identify model architectures and workloads \cite{r25,r26,r31}, while power analysis also recovers cryptographic keys \cite{r27}. Recent recurrent-state studies examine poisoning \cite{r24}, the separation of attention and recurrence \cite{r29}, and record omission \cite{r30}. The confidentiality question here concerns private information retained after its input turn and exposed to a restricted observer when the state is reused. A recurrent-state entry mixes contributions from many tokens, so there is no corresponding per-token entry to evict. Protection of prompt text, responses, and attention \ac{KV} must account for a separate representation whose contents are implicit and whose use can be physically observable.

The disclosure of this content harms the user. One observation of the recurrent state can reveal much of a private history at once: a secret the user supplied, the conditions attached to it, the response rules the application enforces, and personal attributes such as age. This target differs from the model architecture and workload identity recovered by prior GPU side channels \cite{r25,r26}, {[}31{]}: the disclosed content belongs to a user of a known model. Two questions follow. From the software side, does the recurrent state retain recoverable private information under partial and noisy observation, does that information survive later public dialogue, and does it still shape the model's later computation? From the hardware side, does executing a computation that reads the state expose this information physically, to an observer restricted to coarse device signals or to an ordinary account that shares the hardware? We answer these questions with partial-state software probes and physical observers that read neither private histories, responses, nor model tensors.

The adversary knows the model and profiles histories with known labels, then reads one restricted signal: a small noisy subset of recurrent-state coordinates, the power of a device that executes one complete recurrent read, or the whole-device power of a shared GPU while the victim runs in a partition. It succeeds when its predictions for held-out people exceed a baseline fitted on training labels alone. Across two released model families, software probing measures what the state retains and whether it shapes later decisions. Two physical paths then measure whether execution exposes the same information, and a state-isolation experiment attributes the physical signal to the recurrent state rather than to the attention cache or convolution state. Every measured victim was our own workload, and the private histories were constructed for controlled tests or drawn from the released synthetic author corpus SynthPAI \cite{r15}.

\Needspace{8\baselineskip}

Our contributions are:

\begin{enumerate}
\def\labelenumi{\arabic{enumi}.}
\tightlist
\item
  We establish retained recurrent state as a confidentiality boundary for personalized LLM services. Private facts and response rules remain inferable after the conversation moves to public content, and state transplantation connects retained information to later decisions. This gives serving systems a basis for extending prompt protection to state access, reuse, and lifetime (Sections~\ref{sec:5.1}, \ref{sec:5.2}, and~\ref{sec:5.3}).
\item
  We develop a methodology for tracing private information from retained state to execution signals. Shared partial observations measure the breadth of disclosure, matched histories and component interventions identify its source, and separately validated power receivers test physical observability. The distinction lets the community assess a state representation and its implementation under explicit observer capabilities (Section~\ref{sec:3}).
\item
  We demonstrate physical leakage under two deployment-relevant controls: a fixed execution schedule and GPU memory partitioning. Complete-read power and whole-device GPU power disclose private content under their stated access conditions. The results identify separate protection requirements for state management, shared telemetry, and data-dependent switching (Sections~\ref{sec:5.4}, \ref{sec:5.5}, and~\ref{sec:5.6}).
\end{enumerate}

\section{Background and Related Work}\label{sec:2}

\subsection{Inference state and KV-cache growth}\label{sec:2.1}

An autoregressive LLM processes a prompt and generates a response one token at a time. Prompt processing is called prefill; subsequent token generation is called decoding. Both stages produce numerical representations of the preceding context. Retaining these representations lets the model use the history without processing the full prompt again at every decoding step \cite{r2}. In full attention, each token contributes key and value vectors at each attention layer. A later token uses its query vector to weight the stored values according to their keys. The \ac{KV} cache holds these context-derived vectors, separate from the model's learned weights. With model configuration, batch size, and precision fixed, a context of length \(T\) requires \(O(T)\) cache space \cite{r2}. Both a longer input and a longer generated response increase this cost. Here, \ac{KV} cache refers to model data retained for inference. A service may also reuse a matching prompt's \ac{KV} across requests to avoid prefill work. That serving optimization is called prompt caching \cite{r20}. Neither term denotes a processor's hardware cache.

\subsection{Recurrent state in modern LLMs}\label{sec:2.2}

Recurrent layers summarize preceding tokens in fixed-size numerical state. Each update combines that state with the current token representation and produces an output for later layers. Mamba and Mamba-2 use state-space mechanisms \cite{r1,r2}; Gated Delta Networks combine gated forgetting with delta-rule updates \cite{r3}. Modern hybrid LLMs interleave these recurrent mechanisms with attention and, in some designs, \ac{MoE} layers. Representative large releases include Qwen3.5-397B-A17B, which interleaves Gated DeltaNet with gated attention \cite{r34}, and Nemotron 3 Super 120B-A12B, which combines Mamba-2, attention, and \ac{MoE} layers \cite{r35}. In these architectures, the recurrent component has fixed capacity while the attention component retains its own \ac{KV} cache. The two components are distinct session representations with separate update and storage behavior.

This state is a history-dependent latent representation. Its dimensions are fixed by the model, but its values depend on the session. Unlike learned weights, it changes during inference. Unlike an activation discarded after a layer executes, it is retained for subsequent tokens. Information from many tokens can contribute to the same state entries, so an entry need not correspond to a particular word or fact. Carrying this state across input-window boundaries can preserve earlier information; the serving system determines whether to carry it forward or initialize a fresh state. Learned gates can also decay or overwrite information during continued execution \cite{r1,r3}. Private prompt content can enter this state during prefill and remain available when a later public request is processed. For example, a public follow-up to a personal assistant still invokes a computation that reads the state derived from the user's earlier profile. Whether an observer can infer that profile depends on what the state retains and what the computation exposes.

\subsection{Physical observability of inference}\label{sec:2.3}

Computation can expose information without granting the observer access to its tensors. Changing operands changes switching activity and can change measured power even under a fixed instruction sequence \cite{r8}. CSI NN uses electromagnetic measurements to recover neural-network implementation information \cite{r9}, while PowerGAN reconstructs inputs from power measurements of compute-in-memory accelerators \cite{r14}. Data-dependent work provides another path. In an \ac{MoE} layer, a router selects a subset of feedforward subnetworks, called experts, for each token representation. When the dispatcher groups tokens by expert, the selected groups determine which expert computations execute. MoEcho infers prompt attributes from expert-load observations and reconstructs generated content from expert sequences, with end-to-end CPU and GPU channels \cite{r18}. In a recurrent \ac{MoE} model, retained state contributes to the representations supplied to the router. Even fixed public tokens can produce different expert selections when that retained state differs. Numerical operands and selected computations provide the two mechanisms examined in Section~\ref{sec:3.3}.

NVIDIA's \ac{MIG} partitions compute and memory resources among instances \cite{r12}. Behind Bars exploits memory-barrier effects across \ac{MIG} partitions to fingerprint models, infer sequence lengths, and distinguish equal-length \ac{MoE} prompts \cite{r10}. PAGE404 studies shared page-migration counters and driver contention on GH200 \cite{r11}. Energon reads GPU power and temperature through \ac{NVML} without root privilege and identifies the family and hyperparameters of a victim transformer \cite{r25}; in response, NVIDIA recommended disabling telemetry for \ac{MIG} and confidential-computing deployments. Veiled Pathways crosses \ac{MIG} isolation through the shared PCIe link: an attacker in another instance reads \ac{NVML} throughput and identifies which language model a victim runs \cite{r26}. GhostAccess uses unified-memory writes that reach the CPU last-level cache to fingerprint ML and LLM workloads across tenants at over 99\% accuracy \cite{r31}. PLATYPUS recovers AES-NI keys through unprivileged access to processor-package power, aggregating many cipher invocations per sample; its disclosure led Linux to restrict that interface \cite{r27}. Our A100 receiver uses the same access assumption as Energon: an unprivileged process can read GPU telemetry \cite{r25}. It reads parent power, the whole-device power exposed by \ac{NVML}, while the victim runs in a \ac{MIG} instance, without a CUDA context or an allocation in a sibling instance. The target is private content retained by a known model; the receiver aggregates repeated computation on public tokens.

\subsection{Confidentiality of prompts and retained state}\label{sec:2.4}

PromptPeek reconstructs the prompts of other users with the capabilities of an ordinary user by detecting, from changes in serving order under longest-prefix-match scheduling, whether a request reused a victim's stored \ac{KV} cache \cite{r22}. \ac{KV}-Cloak shows that an inference provider holding the \ac{KV} cache can recover the user's prompt, since \ac{KV} entries correspond element-wise to input tokens and the cache of a single layer suffices \cite{r23}. PRSA uses limited input-output examples to infer prompt intent and construct prompts that reproduce its functionality \cite{r19}. Prompt-cache audits use response timing to detect reuse of matching prefixes across users \cite{r20}. Song and Raghunathan study attribute, membership, and inversion leakage from embeddings \cite{r6}; Morris et al.~reconstruct source text from embeddings \cite{r7}. Staab et al.~show that pretrained LLMs infer personal attributes such as age and location from user-written text, for an adversary who holds the text or operates the chatbot that receives it \cite{r28}; our observers receive neither the text nor the responses. Qu et al.~reconstruct prompts from activations received by a participant in collaborative LLM inference \cite{r21}; that participant receives activations of the request it processes, whereas our observers see partial, noisy views of state retained from an earlier private turn.

Recurrent-state research also distinguishes several security and behavioral properties. Hidden-state poisoning shows that short input phrases overwrite information in Mamba-based models, targeting the integrity of retained state \cite{r24}. A mechanistic split-prefill study swaps one context's attention cache with another's recurrent state and finds that recurrence governs language and persona \cite{r29}. Record-omission work in delta attention finds that masking removes the exact answer while a query containing a known candidate can still infer it \cite{r30}. We use component intervention to test the recurrent contribution to later decisions and study recovery of unknown private content from partial state and power. The target includes explicit facts, application response rules, and attributes associated with an author's text; an attribute label need not itself appear in that text. These targets concern what a restricted observer can learn about a retained history, including when the measured request contains only public tokens.

\section{Method}\label{sec:3}

\subsection{Overview and threat model}\label{sec:3.1}

We trace confidentiality through three relationships: private history to recurrent state, retained state to later computation, and computation to an observer's signal. Software probing first tests whether restricted state observations predict private content. Applying a frozen predictor after public dialogue tests persistence; transplanting state between matched histories tests whether that retained information contributes to later decisions (Section~\ref{sec:3.2}). Physical observation then tests two consequences of using the state: data-dependent power during a complete recurrent read and changes in expert execution during public requests (Section~\ref{sec:3.3}). The first path holds the execution schedule fixed. The second isolates the recurrent contribution before testing a service that retains complete session caches. These comparisons distinguish information in a representation from leakage through an implementation.

The victim's prompt history determines its recurrent state under a fixed model and extraction procedure. The target is a private fact, an assigned response rule, or a labeled personal attribute inferred from the history. Controlled prompts supply facts and rules. Author-profiling corpora supply text to the victim; profile labels are kept outside model input. The adversary knows the model, profiles histories with known labels, and fits a task head before observing test people. An inference counts as disclosure when it exceeds the training-label baseline on held-out people under the stated statistical test.

\begin{table*}[t]
\centering
\caption{Observation access and evaluation roles}
\label{tab:access}

\small
\begin{tabularx}{\textwidth}{>{\raggedright\arraybackslash}X>{\raggedright\arraybackslash}X>{\raggedright\arraybackslash}X>{\raggedright\arraybackslash}X}
\toprule
Path & Observer access & Controlled execution & Predicted target \\
\midrule
Software observation & Selected recurrent coordinates & State extraction from released models & Private facts, rules, and author attributes \\
Complete-read power & Labeled profiling and local CW305 power traces & Repeated complete recurrent reads; matched and fixed-coefficient controls & Condition and vault word \\
A100 parent power & Host parent power; opaque session handles in the service run & Repeated public-token target/anchor batches and calibration & Age \\
\bottomrule
\end{tabularx}
\end{table*}

Table \ref{tab:access} separates the observer capabilities. The software observer receives selected recurrent coordinates, modeling state exposure during caching, offloading, migration, or sharing. The complete-read observer has local physical access to power traces from the device executing the read. The A100 observer is an unprivileged host process with access to whole-device power and no CUDA context. In the natural-session service, this process also holds append-only handles to target and public-anchor sessions. The service grants these handles under its host policy; the attacker does not obtain another tenant's handles by exploiting the model. It receives no private history, test label, cache tensor, response, or route truth. Both GPU variants require repeated public requests, public anchors, and same-session calibration. A public anchor is a reference history available for calibration, not a test author's private history.

The attack has three stages. Profiling uses labeled histories from other people to select an observation and fit its receiver and task head. Acquisition records the permitted state observation or physical signal for a held-out person. Inference applies the frozen receiver and task head, with test labels reserved for scoring. A physical receiver can estimate an intermediate observation for a task head or predict the target directly from a waveform. The GPU receiver estimates routing distances to public anchors; the complete-read receiver predicts private facts from traces. Separate software and physical tests expose errors introduced by acquisition instead of treating an informative internal representation as evidence of a working physical channel.

\subsection{Software probing}\label{sec:3.2}

\subsubsection{Partial-state inference}\label{sec:3.2.1}

A partial-state observation limits both the coordinates and the precision available to the observer. Profiling selects coordinates or fixes a label-blind sample, and a task head learns to predict a private label from that observation. Varying the coordinate budget and observation noise tests which information remains accessible under progressively restricted access. Raw-value and Hamming-weight observations distinguish a state representation's information content from the precision needed to recover it. The Hamming-weight interface is an observation model; each physical receiver is separately evaluated on measured power. Section~\ref{sec:4.3} specifies the encoding and classifiers.

A common observation can support several task heads. In the multi-task experiment, the condition, vault, and command heads all receive the same coordinate vector. A separate experiment freezes the acquisition budget using a subset of tasks before exposing labels for later tasks. This tests whether task-specific heads can use an already chosen observation, with acquisition selection and subsequent prediction kept as distinct steps.

\subsubsection{Retention and causal contribution}\label{sec:3.2.2}

To measure retention, we fit a receiver at the initial checkpoint and apply it unchanged after later public dialogue. Matched histories preserve public text while changing the private condition or response rules. Predictions are compared on the same people, so public dialogue alone cannot explain a difference. To measure contribution to later computation, we transplant recurrent state between matched histories before the future query. Attention \ac{KV} remains from the source history. The future query and response are processed only after transplantation; convolution state is a separate intervention arm. The decision margin is the donor marker logit minus the source marker logit. A change in this margin measures the recurrent state's causal contribution to the later answer. Same-source copying is the control.

For the pre-query intervention, the snapshot boundary is taken before the entire future user turn, using the rendered conversation to identify that boundary. This keeps both the query tokens and response tokens downstream of the intervention. Each source/donor pair has the same public context and differs in its assigned VERIFY rule. Replacing the recurrent matrices while retaining source \ac{KV} asks whether the recurrent component changes the donor-versus-source score on its own. We compare all recurrent matrices, the final six recurrent layers, and recurrent plus convolution state. These interventions distinguish a distributed effect from one reproduced by a later subset and test whether convolution state accounts for the result. The observational and interventional experiments answer separate questions: a task head measures accessible information, whereas transplantation measures its effect on a later model computation.

\subsection{Physical observability}\label{sec:3.3}

The physical paths test whether later use of retained state exposes private information without tensor access. Complete-read power holds control flow fixed to test leakage from numerical operands. Routing-dependent power observes changes in the work selected by a recurrent \ac{MoE} model under public tokens. Each path has its own profiling data, control input, and receiver; neither assumes that successful software inference guarantees a measurable physical signal.

\subsubsection{Complete-read power}\label{sec:3.3.1}

To test whether state values leak even under a fixed execution schedule, we implement complete reads of a model-derived recurrent operator. All terms of each selected output are executed. A matched arm preserves the state and coefficients from one model invocation; a fixed-coefficient arm varies only the confidential state across people. Their power measurements separate leakage from confidential operands from variation in the read schedule.

For one Gated DeltaNet head, write the incoming state as \(S\in\mathbb{R}^{128\times128}\), the normalized and implementation-scaled query as \(q\), the normalized key as \(k\), the value as \(v\), the decay as \(\alpha=\exp(g)\), and the update gate as \(\beta\). Equation \eqref{eq:state-update} gives the relevant update and read:

\begin{equation}
\label{eq:state-update}
S'=\alpha(I-\beta kk^T)S+\beta kv^T,
\qquad o^T=q^TS'.
\end{equation}

The old-state contribution to output column \(r\) follows in Equation \eqref{eq:old-state}:

\begin{equation}
\label{eq:old-state}
o_{\mathrm{old},r}=c^TS[:,r],
\qquad c=\alpha\big(q-\beta(q^Tk)k\big).
\end{equation}

We capture \(S,q,k,g,\beta\) from the same Transformers forward invocation. The FPGA executes the selected old-state term \(c^TS[:,r]\) for complete output columns. Core profiles select the columns; every selected column includes all 128 terms. State values use E4M3FN, coefficients use core-scaled int16, and products accumulate in 48 bits. This preserves the specified read operator and its state dependence under quantization. The victim-matched arm uses each person's own state and coefficients. Its sham retains the coefficients and schedule while substituting public state at the multiplier. Since real state can still reach an input register in that sham, a second arm holds one coefficient vector fixed across people and substitutes public bytes at the external input. The person whose coefficients are reused, the donor, is chosen from core profiles by four-fold validation and then fixed for the independent test population.

The fixed-coefficient contrast varies the confidential state while holding coefficients, addresses, read order, and control flow fixed. Randomized acquisition reduces scheduling confounds. The control state depends on public start inputs and the previous control state, so equal public inputs give equal chunk indices and completion schedules by induction. The remaining data-dependent switching is the measured signal. Arithmetic simulation and a finite schedule-equivalence check validate the implemented RTL. For attribution, equal schedules are not enough: the coefficient tensor must also be constant across people, and the public-input control must keep confidential bytes out of the loading path. The fixed-coefficient arm satisfies both, and randomized record order prevents a fixed label ordering from defining the schedule, so the actual-versus-control comparison measures variation introduced by processing the confidential state.

The receiver is trained on paired actual-minus-sham traces or on actual traces alone, with sham predictions providing an attribution comparison. Column selection preserves the whole read: every selected output accumulates all state entries in that column with the extracted coefficient vector, and an independent integer reference checks the completed output. Section~\ref{sec:4.4} gives the chunking, repetition counts, and acquisition protocol.

\subsubsection{Routing-dependent power}\label{sec:3.3.2}

The routing path tests whether retained private context changes execution during a later public request. A target session is combined with copies of a public-anchor session, and an all-anchor batch provides a control. In the state-isolation variant, only the target's recurrent matrices come from private history; attention \ac{KV} and convolution state come from a common public prefix. In the natural-session variant, the service retains the complete cache produced by ordinary prefill. The first identifies a recurrent contribution, while the second tests the complete request-to-power path available through the service interface.

We instantiate this path with Granite-4.0-H-Tiny \cite{r16}. Routing work counts the additional expert groups executed when a target is combined with a public anchor. For target session \(i\), anchor session \(j\), token \(t\), and \ac{MoE} layer \(\ell\), let \(R_{i\ell t}\) and \(A_{j\ell t}\) denote the six experts selected by their lanes within the mixed batch. Equation \eqref{eq:route-work} defines batch routing work:

\begin{equation}
\label{eq:route-work}
Z_{ij}=\sum_{\ell=1}^{40}\sum_{t=1}^{8}
       \left|R_{i\ell t}\setminus A_{j\ell t}\right|.
\end{equation}

The batch contains 15 identical anchor lanes and one target lane. Replacing the target with another anchor gives an all-anchor sham that executes \(40\times8\times6=1{,}920\) nonempty expert linear groups. The target batch executes exactly \(1{,}920+Z_{ij}\) groups. The difference arises from expert selection during the released Granite and Transformers forward path. We count groups in an instrumented run before acquisition and use the stock dispatcher during the timed physical probe.

At each layer and token, the anchor lanes activate the same six experts. Adding the target lane leaves those experts active and adds exactly the target experts absent from the anchor set. The union contains \(6+|R_{i\ell t}\setminus A_{j\ell t}|\) experts. Summing this identity over the fixed layers and tokens gives the total group count. This is an identity about the released dispatcher's nonempty expert groups; the time required by those groups remains an empirical relationship. The all-anchor sham preserves token identities, batch shape, cache preparation, and repetition count. In the state-isolation run, it replaces the target recurrent matrices with public anchor state. In the natural-session run, it replaces the target lane's complete cache with an independently prefetched anchor cache.

A second variant uses naturally prefetched complete session caches. The service concatenates independent attention \ac{KV}, convolution, and recurrent caches from 15 public-anchor sessions and one target session. The union identity in Equation \eqref{eq:route-work} still describes expert groups. This variant tests whether one host client can cause and observe a history-dependent computation. Its routing signal can include effects of attention \ac{KV} and convolution state. The state-isolation variant identifies the recurrent contribution under the same public-token mechanism.

Passivity describes how the victim produces the signal, while the client only chooses public requests. The A100 victim does not evaluate an age predicate or encode a secret symbol; its expert dispatch follows the released model's forward computation on fixed public tokens. In both contracts the receiver obtains power from the victim's computation rather than a sender-controlled encoding.

A profiling run extracts singleton routing distances and fits an age head using four public anchors. Each physical run then executes target/anchor batches and all-anchor shams. The state-isolation run uses a separate host power observer; the natural-session run has the public-request client sample parent power under the same unprivileged account.

The observer estimates execution duration from a power transition. A candidate change point splits a record into active and idle segments; its objective is the sum of squared deviations from each segment's mean. The search window and receiver parameters are fixed before independent validation. Subtracting the sham change point from the target change point gives one paired shift for each public anchor. An ordinary least-squares map fitted on calibration data converts this shift to the singleton distance expected by the frozen age head, with estimates clipped to the calibration range. The four distances form the task observation. Neither these maps nor the age head use terminal labels. The receiver uses transition timing, not the power level as a state value.

Calibration is needed because the age head uses singleton executions, while physical routing work comes from a mixed target/anchor batch whose routes can differ. Independent mechanism validation tests the work-to-duration and duration-to-power relationships and their composition, and the terminal task test evaluates the complete calibrated chain against the same head applied to clean software distances.

\section{Experimental Setup}\label{sec:4}

\subsection{Models}\label{sec:4.1}

We instantiated state-space and gated-delta recurrence using the released Granite-4.0-H-350M and Qwen3.5-2B implementations \cite{r4,r5}. Our extracted recurrent states contained \ResultGraniteStateElements{} and \ResultQwenStateElements{} elements. Qwen used 18 Gated DeltaNet layers and six full-attention layers. The routing experiments used Granite-4.0-H-Tiny \cite{r16}. These implementations support direct state extraction, component intervention, and execution measurements of the mechanisms defined in Section~\ref{sec:3}. Observation vectors excluded attention \ac{KV}, which was controlled separately in the interventions.

\subsection{Benchmarks and populations}\label{sec:4.2}

The Granite corpus tests how observation budget and noise affect recovery of one private fact and one response rule. The Qwen corpus adds independent facts and rules to test how much a single observation can recover before their queries. The Granite corpus contained 128 synthetic people, each with two sessions, a four-way condition, and a four-way response rule. Training used 96 people and wording templates 0--3; testing used the other 32 people and templates 4--7. Two runs regenerated both the corpus and coordinate selection. Qwen used 384 training and 128 test people per seed, with disjoint wording families. Its histories combined an early condition, an independently paired vault word, public contextual fields, and independently assigned four-way PING, STATUS, and VERIFY responses. Condition/vault combinations were balanced in each split. Granite's snapshot followed the public PING query and preceded its response. Qwen's snapshot followed a fixed public READY interaction and preceded every private-fact and command query, keeping those queries out of the shared observation. Both corpora hold out people and wording families; the different targets, populations, and snapshot boundaries make each accuracy a result for its own task and controls, rather than a ranking of the two models.

For tasks introduced after acquisition selection, a third independent Qwen corpus used 384 core, 128 development, and 128 test people. Condition, vault, and PING labels selected among five nested budgets that doubled from 512 to 8,192 coordinates. STATUS and VERIFY labels were supplied only after the acquisition choice was frozen. Their heads were then fitted on the allowed training population and evaluated on untouched people.

SynthPAI \cite{r15} supplied all 300 synthetic authors and all \ResultSynthComments{} comments. Author IDs and profile-label metadata were excluded from victim input. A label-independent hash assigned authors to five folds of 60. Software results used author-level out-of-fold predictions. A100 development and head fitting used folds 1--3; both terminal tests covered all 60 authors in fold 4. Each run selected 24 calibration authors from folds 1--3 without labels to cover its four-anchor distance space. Each clean fold-4 result was known before its physical acquisition; the physical tests measured preservation of those frozen software predictions.

The longitudinal experiment followed 128 test people through 0, 4, 12, and 24 public turns. The pre-query intervention used 32 profiles across two wording families, four history lengths, and four private conditions, giving 384 ordered source/donor pairs. Uniform four-class guessing is 25\%; response-rule accuracy refers to the assigned response rule. Actual model responses are scored separately when behavior is the endpoint.

\subsection{Software observation protocols}\label{sec:4.3}

For sparse software observations, selected coordinates are stored in 16-bit floating point. We quantize them to E4M3FN, an eight-bit floating-point format with four exponent bits and three mantissa bits. The deterministic encoder saturates out-of-range values and resolves ties toward lower magnitude. Each coordinate is then reduced to the Hamming weight of its eight-bit code, an integer from 0 to 8. A fixed coordinate permutation defines nested budgets. Training-only coordinate selection ranks between-class versus within-class variation. A standardized ridge classifier with ridge parameter 1 predicts task labels using training-derived scales. Noise experiments add Gaussian perturbations in held-out Hamming-weight units while leaving the task head trained on noiseless profiles. Repeated noise draws measure sensitivity to observation error on the same people. They are averaged within the reported endpoint and do not enlarge the held-out population. SynthPAI also evaluates sampled raw recurrent values. Each result specifies which representation it uses. The Hamming-weight interface is a reproducible way to test how much numerical precision a task needs; it is part of our observation model, not a claim that the released model stores its state in E4M3FN. The physical experiments instead obtain their inputs from waveforms or parent power and validate those receivers separately.

\subsection{Complete-read acquisition}\label{sec:4.4}

CW305 measurements used ChipWhisperer 6.0.0 \cite{r13} and a Vivado 2023.2 implementation, with a 10 MHz FPGA clock, and 40 MS/s acquisition. One clear and eight \ac{MAC} cycles produced a 36-sample period; receivers used the first 2,304 of 4,096 retained samples. The implementation met a 100 MHz timing constraint. Profiling used 384 core and 128 development people. Two disjoint, balanced groups of 64 supplied the victim-matched and fixed-coefficient tests. Profile, bridge, and test sessions used separate processes and forced board reprogramming. Four independently averaged test blocks measured repeatability.

The FPGA serialized each complete output into 16 chunks using one DSP slice and a 48-bit accumulator. The implemented design used \ResultSynthesisLuts{} \acp{LUT} and \ResultSynthesisRegisters{} registers. Profile and bridge records averaged ten external complete executions; each test block averaged 25. Record order was randomized within each acquisition. A bridge session remeasured 32 known profile people after board reprogramming, while the terminal sessions used previously unseen people. Bridge agreement and unseen-person accuracy measured different forms of transfer. Each eight-term chunk repeated 64 times for acquisition and committed once; this internal repetition is separate from the external complete-execution averages.

The victim-matched acquisition measured eight selected output slots. Development fixed slot 2 for condition and slots 2--3 for vault. The fixed-coefficient arm retained those two slots and selected its coefficient donor using only core profiles. A four-fold validation rule maximized the smaller of the two task accuracies, then their sum, with donor index breaking remaining ties. The selected tensor was reused byte-for-byte in profile, bridge, test, actual, and sham records.

\subsection{GPU power acquisition}\label{sec:4.5}

The passive endpoint used an A100 80GB PCIe with a victim in one \ac{MIG} instance holding three compute slices and 40 GB of memory. Acquisition records identify driver \ResultAgeDriver{} and PyTorch \ResultAgeTorch{}, built for CUDA \ResultAgeCuda{}. The device power limit was \ResultAgePowerLimit{} W. The unused sibling instance was removed before acquisition, persistence mode was disabled, and an idle 12 GiB allocation inside the victim instance kept other jobs from starting there. A process running as a different unprivileged user created no CUDA context and read parent power through \ac{NVML} every 20 ms. The harness prepared state copies before each timed probe. Acquisition used repeated target/anchor batches, paired shams, and a frozen schedule. This was a controlled host measurement of a \ac{MIG}-resident victim. In a separate access test on the same A100, a non-root process inside a container allocated only an existing \ac{MIG} instance with one compute slice and 10 GB of memory could read parent-device power through \ac{NVML}. The test established telemetry access for a \ac{MIG} tenant on this host; the state-isolation terminal traces were acquired by the host observer.

An independent mechanism panel used 32 targets excluded from development and calibration. A label-blind hash assigned eight targets to each anchor. Each pair processed eight public tokens on 32 cache copies, with the change-point search fixed to 25--35 s of a 36 s record. The terminal four-anchor acquisition used 28 copies and 40 s slots for all 24 calibration and 60 test authors. GPU timing and instrumented routing truth were available to offline analysis, not to the measured observer.

Each terminal record contained 1,800 power samples over 36 seconds, with the transition search fixed to 18--35 seconds; the recorded rule carried forward the \ResultAgeInvalidSamples{} invalid power reads. The 40-second slot included the interval reserved for a record. State-copy preparation finished before probe timing. This kept preparation outside the work measured by the change-point receiver. The terminal probe used 28 independent cache copies, so each copy began from the intended retained state before the fixed public-token sequence was processed. The calibration panel was chosen by farthest-point coverage of standardized four-anchor distances from training folds. This selection used geometry rather than age labels. Calibration measured all anchors for each selected author, allowing one map per anchor. Every terminal author was likewise measured against every anchor. This state-isolation endpoint covers the full held-out fold.

The natural-session endpoint used the existing 3g.40gb victim instance with an existing 1g.10gb sibling. A controlled service retained complete caches for \ResultNaturalAgeSessions{} sessions after \ResultNaturalAgeHistoryTokens{}-token prefills. One unprivileged host account submitted fixed public continuations through opaque handles and sampled parent power without a CUDA context. Its UID could not read the private corpus or handle map, and the service returned no model response. Each batch had \ResultNaturalAgeBatchLanes{} lanes and \ResultNaturalAgeProbeRepeats{} copies of \ResultNaturalAgePublicTokens{} public tokens. A \ResultNaturalAgeSlotSeconds{}-second slot yielded \ResultNaturalAgeSamplesPerRecord{} parent-power samples. The receiver, request schedule, and age head were frozen before acquisition.

\subsection{Statistical analysis and data use}\label{sec:4.6}

Core and development data selected coordinates, quantization scales, output columns, and receivers. These choices were frozen before their test acquisitions. Test labels were used for scoring. The person or author was the unit of generalization; repeated blocks and waveform samples did not increase that population size. Physical transfer tests used separately programmed bridge sessions and new test people. SynthPAI used five-fold author-level prediction with fold-training majority or median baselines. Age and all eight categorical fields formed one Holm-corrected family. Both A100 age heads were fixed before acquisition; age \ac{MAE} was each terminal endpoint. Training-only majority and median baselines account for label imbalance; clean coordinates provide the task reference before physical acquisition; public-state and sham-only inputs test attribution. Each baseline uses the same held-out people as its prediction. The first A100 terminal labels were opened after predictions were fixed. The natural-session acquisition reused that held-out fold with a receiver frozen before its new measurements.

Controlled tasks used binomial references and Holm correction over stated endpoint families. Real-versus-control comparisons used paired tests. Conditional permutations tested vault recovery within true-condition strata. Intervention intervals clustered by profile. Routing correlations used anchor-stratified permutations; age improvement used a paired sign-flip test. A supplementary author bootstrap summarizes uncertainty in the state-isolation age result: 50,000 paired resamples of the 60 authors, with a fixed seed and percentile 95\% intervals. It holds predictions, calibration, and the fitted head fixed; it measures variation across authors within this acquisition, not across independently calibrated sessions.

Custom histories and SynthPAI profiles are synthetic. Hardware and cloud acquisitions target our own victim workloads. Other tenants, when present in exploratory runs, are background activity rather than data sources or targets. The experimental record retains model-extraction manifests, splits, receiver choices, bitstreams, raw traces, and predictions. Replay scripts compare retained records and recompute task results and statistical tests; Appendix~\ref{app:A} lists the manifests and replay entry points. The reproducibility record distinguishes the target snapshot from the current environment.

\section{Results}\label{sec:5}

\subsection{Private information from partial observations}\label{sec:5.1}

\begin{figure}[t]
\centering
\providecommand{\bandline}[3]{%
  \addplot[name path=#1lo, draw=none, forget plot] table[x=budget, y expr=100*\thisrow{#1lo}] {figdata/granite_budget.dat};
  \addplot[name path=#1hi, draw=none, forget plot] table[x=budget, y expr=100*\thisrow{#1hi}] {figdata/granite_budget.dat};
  \addplot[#2, fill opacity=0.18, draw=none, forget plot] fill between[of=#1lo and #1hi];
  \addplot[#2, mark=#3, mark size=1.4pt] table[x=budget, y expr=100*\thisrow{#1}] {figdata/granite_budget.dat};}
\subfloat[Private condition]{\begin{minipage}{.49\linewidth}\centering
\begin{tikzpicture}
\begin{axis}[paperplot, width=\linewidth, height=4.1cm,
  xmode=log, log basis x=2, xmin=8, xmax=8192,
  xtick={8,64,512,4096}, xticklabels={8,64,512,4096},
  ymin=0, ymax=105, ytick={0,25,50,75,100},
  xlabel={observed coordinates},
ylabel={accuracy (\%)},  title style={font=\sffamily\footnotesize\bfseries, yshift=-1mm},
  legend to name=budgetlegend, legend columns=4, legend style={/tikz/every even column/.append style={column sep=3pt}}, legend cell align=left]

\bandline{cond0}{tolblue}{*}
\bandline{cond1}{tolblue!70!black}{square*}
\bandline{cond2}{tolgrey}{triangle*}
\addplot[negc, dash pattern=on 2pt off 1.5pt, line width=0.6pt] coordinates {(8,25) (8192,25)};
\legend{noise sd 0, noise sd 1, noise sd 2, chance}
\end{axis}
\end{tikzpicture}
\end{minipage}}\hfill
\subfloat[Response rule]{\begin{minipage}{.49\linewidth}\centering
\begin{tikzpicture}
\begin{axis}[paperplot, width=\linewidth, height=4.1cm,
  xmode=log, log basis x=2, xmin=8, xmax=8192,
  xtick={8,64,512,4096}, xticklabels={8,64,512,4096},
  ymin=0, ymax=105, ytick={0,25,50,75,100},
  xlabel={observed coordinates},
title style={font=\sffamily\footnotesize\bfseries, yshift=-1mm}]

\bandline{rule0}{tolblue}{*}
\bandline{rule1}{tolblue!70!black}{square*}
\bandline{rule2}{tolgrey}{triangle*}
\addplot[negc, dash pattern=on 2pt off 1.5pt, line width=0.6pt] coordinates {(8,25) (8192,25)};
\end{axis}
\end{tikzpicture}
\end{minipage}}
\par\smallskip\ref{budgetlegend}
\caption{Granite accuracy on 64 held-out records from 32 people as a function of the number of randomly observed recurrent coordinates and the standard deviation of Gaussian noise added to their Hamming weights. Lines are the mean of two independent corpus and coordinate seeds; shading spans the two seeds.}
\label{fig:2}
\end{figure}
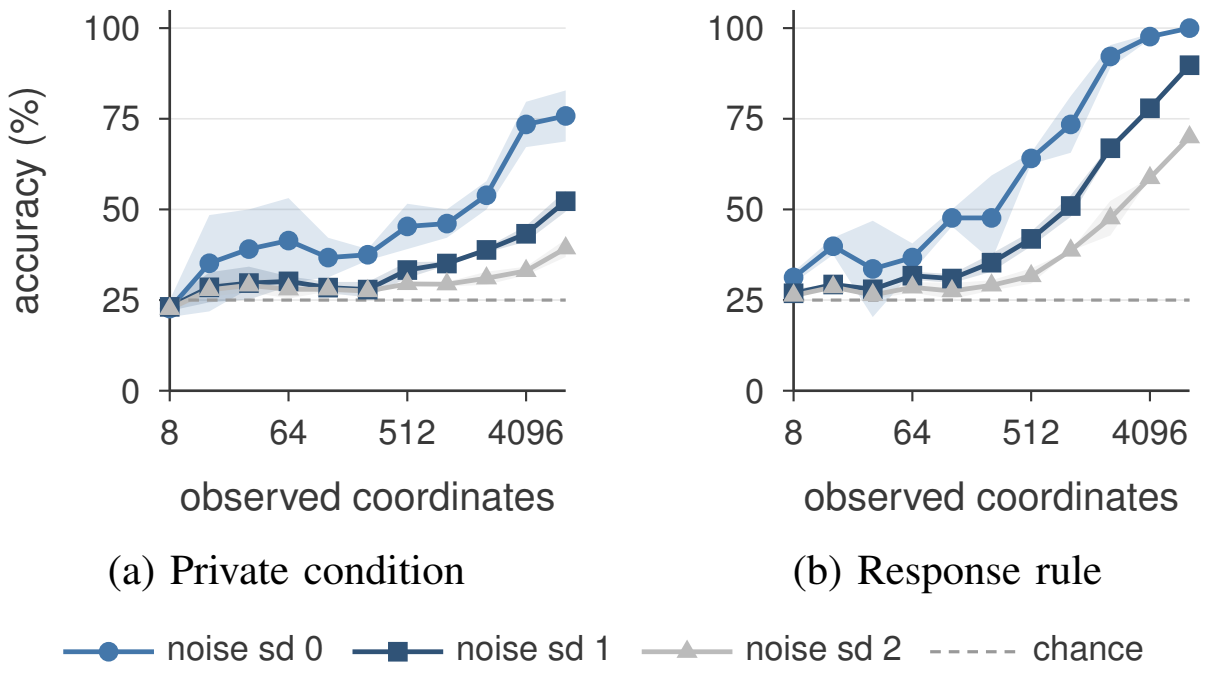

The Granite experiment varied observation budget and numerical precision on the same held-out records. An 8,192-coordinate random sample covered \ResultGraniteFraction{}\% of recurrent state. Without added noise, condition accuracy was \ResultGraniteOneCondition{}\% and \ResultGraniteTwoCondition{}\% in the two runs; response-rule accuracy was \ResultGraniteOneRule{}\% in both. The model followed the assigned response rules on \ResultGraniteOneCompliance{} and \ResultGraniteTwoCompliance{} records in the respective corpora, distinguishing the rule label from the model's observed compliance. At Hamming-weight noise standard deviation 1, condition accuracy was \ResultNoiseOneConditionMin{}\%--\ResultNoiseOneConditionMax{}\% and rule accuracy was \ResultNoiseOneRuleMin{}\%--\ResultNoiseOneRuleMax{}\%. At standard deviation 2, the respective ranges were \ResultNoiseTwoConditionMin{}\%--\ResultNoiseTwoConditionMax{}\% and \ResultNoiseTwoRuleMin{}\%--\ResultNoiseTwoRuleMax{}\%, against a 25\% guessing reference. Figure~\ref{fig:2} reports the full budget and noise grid. The two runs regenerated both the corpus and coordinates, so the trend covers both sources of variation.

Training-selected subsets of 32 coordinates achieved \ResultGraniteOneSelected{}\% condition accuracy in both runs. With about 300 coordinates observed per layer, the final two recurrent layers yielded \ResultGraniteLateOne{}\%--\ResultGraniteLateTwo{}\% rule accuracy and contained \ResultGraniteOneLateSelected{} and \ResultGraniteTwoLateSelected{} of the selected rule coordinates. The first run's 32 selected condition coordinates spanned \ResultGraniteConditionLayers{} layers. Task-directed selection found different spatial patterns for a stored fact and a response rule, so the location of the observed coordinates, in addition to their number, determined which information was accessible.

On Qwen, the same 8,192-coordinate observation was supplied to every task head, covering \ResultQwenFraction{}\% of recurrent state (Table \ref{tab:qwen}). It supported both private facts and three separately assigned response rules on unseen people. The response rules are assigned instructions, while the private condition and vault word are facts supplied in the history. Joint scoring requires both private facts, or all three response rules, to be correct for the same person; it tests whether one observation discloses multiple components of a private history without increasing the observation budget.

\begin{table*}[t]
\centering
\caption{Qwen inference from a shared partial observation}
\label{tab:qwen}
\small
\begin{tabular}{lrrr}
\toprule
Qwen task & Seed 1, 128 unseen people & Seed 2, 128 unseen people & Chance reference \\
\midrule
Private condition & \ResultQwenOneCondition{}\% & \ResultQwenTwoCondition{}\% & 25\% \\
Private vault word & \ResultQwenOneVault{}\% & \ResultQwenTwoVault{}\% & 25\% \\
PING rule & \ResultQwenOnePing{}\% & \ResultQwenTwoPing{}\% & 25\% \\
STATUS rule & \ResultQwenOneStatus{}\% & \ResultQwenTwoStatus{}\% & 25\% \\
VERIFY rule & \ResultQwenOneVerify{}\% & \ResultQwenTwoVerify{}\% & 25\% \\
Both private facts & \ResultQwenOneBothPrivateFacts{}\% & \ResultQwenTwoBothPrivateFacts{}\% & 6.25\% \\
All three rules & \ResultQwenOneAllThreeFuture{}\% & \ResultQwenTwoAllThreeFuture{}\% & 1.5625\% \\
\bottomrule
\end{tabular}
\end{table*}

The independent acquisition-freeze experiment selected the 8,192-coordinate budget before STATUS and VERIFY labels were supplied. Its later heads correctly predicted STATUS for \ResultLateStatusCount{} people and VERIFY for \ResultLateVerifyCount{}, with Holm-adjusted \(p=\ResultLateStatusP{}\) and \(\ResultLateVerifyP{}\). The model followed those rules on \ResultLateStatusBehavior{} and \ResultLateVerifyBehavior{} of the same people. The result establishes that the chosen observation contained information usable by tasks excluded from its selection. Each correct prediction here is a private fact or an enforced rule taken from a person whose conversation the adversary never saw.

\subsection{Retention and future computation}\label{sec:5.2}

\begin{figure}[t]
\centering
\subfloat[Private condition]{\begin{minipage}{.49\linewidth}\centering
\begin{tikzpicture}
\begin{axis}[paperplot, width=\linewidth, height=4.1cm,
  xmin=-1.5, xmax=25.5, xtick={0,4,12,24},
  ymin=0, ymax=100, ytick={0,25,50,75,100},
  xlabel={later public turns},
  title style={font=\sffamily\footnotesize\bfseries, yshift=-1mm},
ylabel={accuracy (\%)},
  legend to name=longitudinallegend, legend columns=3, legend style={/tikz/every even column/.append style={column sep=3pt}}, legend cell align=left]

\addplot[obsc, mark=*, mark size=1.5pt] table[x=turns, y expr=100*\thisrow{condfull}] {figdata/longitudinal.dat};
\addplot[tolgrey!80!black, mark=o, mark size=1.5pt] table[x=turns, y expr=100*\thisrow{condctl}] {figdata/longitudinal.dat};
\addplot[negc, dash pattern=on 2pt off 1.5pt, line width=0.6pt] coordinates {(-1.5,25) (25.5,25)};
\legend{original history, matched control, chance}
\end{axis}
\end{tikzpicture}
\end{minipage}}\hfill
\subfloat[Response rule]{\begin{minipage}{.49\linewidth}\centering
\begin{tikzpicture}
\begin{axis}[paperplot, width=\linewidth, height=4.1cm,
  xmin=-1.5, xmax=25.5, xtick={0,4,12,24},
  ymin=0, ymax=100, ytick={0,25,50,75,100},
  xlabel={later public turns},
  title style={font=\sffamily\footnotesize\bfseries, yshift=-1mm},
]

\addplot[obsc, mark=*, mark size=1.5pt] table[x=turns, y expr=100*\thisrow{polfull}] {figdata/longitudinal.dat};
\addplot[tolgrey!80!black, mark=o, mark size=1.5pt] table[x=turns, y expr=100*\thisrow{polctl}] {figdata/longitudinal.dat};
\addplot[negc, dash pattern=on 2pt off 1.5pt, line width=0.6pt] coordinates {(-1.5,25) (25.5,25)};
\end{axis}
\end{tikzpicture}
\end{minipage}}
\par\smallskip\ref{longitudinallegend}
\caption{Persistence of private information. A receiver fitted once before any later turn is applied unchanged after 4, 12, and 24 public turns to 128 held-out people (original history) and to matched controls that keep the public trajectory and token count but replace the private condition.}
\label{fig:3}
\end{figure}
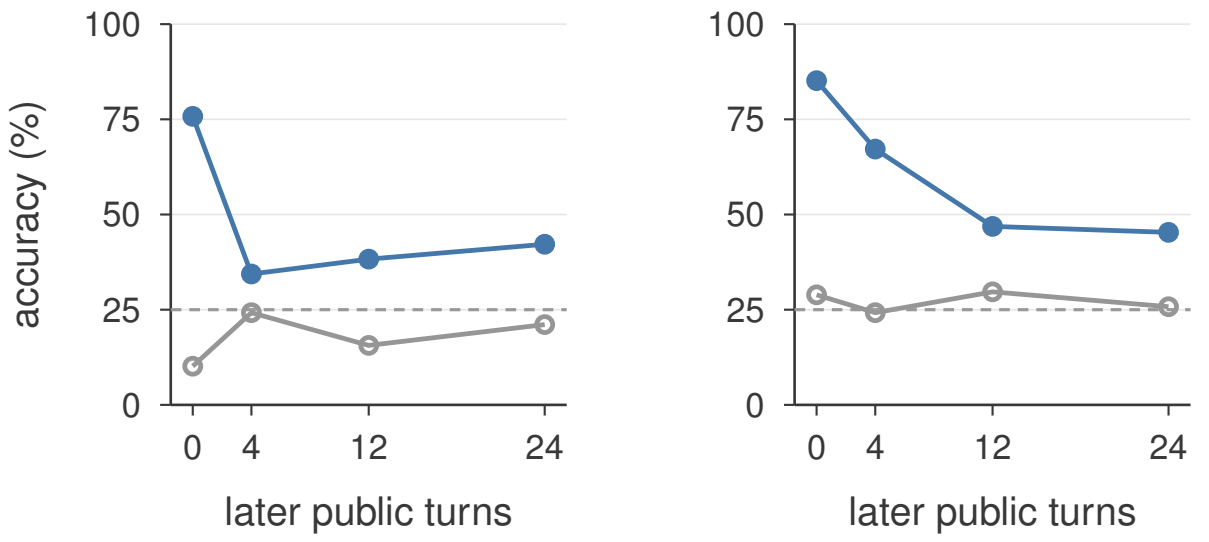

The initial-checkpoint Qwen receiver retained predictive information after public dialogue (Figure~\ref{fig:3}). At 24 turns, condition accuracy was \ResultRetentionConditionFull{}\% against \ResultRetentionConditionControl{}\% on matched controls, and response-rule accuracy was \ResultRetentionPolicyFull{}\% against \ResultRetentionPolicyControl{}\%. Corrected paired tests gave \(p=\ResultRetentionConditionP{}\) and \(\ResultRetentionPolicyP{}\). The histories contained \ResultRetentionTokensMin{}--\ResultRetentionTokensMax{} tokens. The response-rule endpoint measures the assigned rule label; the model's subsequent compliance is a separate behavioral quantity. At each checkpoint, the control kept the public alias, conversation, and exact token count while replacing the private condition and removing the original response rules, and was scored against the original secret labels. This matched comparison tests whether the public trajectory alone explains the output, with the unchanged initial head applied to both trajectories.

The checkpoint curve is not a simple monotone decay: condition accuracy fell after the first public turns yet stayed above its matched control at the final checkpoint, and the response-rule endpoint kept an advantage after the original instruction had left the current topic. The measured property is persistence of inferable historical content under the specified public dialogue.

Pre-query state transplantation shifted the donor-directed decision margin on every one of the \ResultInterventionPositivePairs{} source/donor pairs while attention \ac{KV} remained fixed (Table \ref{tab:intervention}). The mean shift for all recurrent matrices was \ResultInterventionRecurrentMean{} logits. All \ResultInterventionPositiveProfiles{} profile-level effects were positive. One million profile-level sign-flip draws gave Monte Carlo \(p\leq\ResultInterventionPBound{}\) for each recurrent intervention. The mean effect was \ResultInterventionInitial{} logits at zero public turns and \ResultInterventionFinal{} after 24 turns, connecting information retention to later computation across the tested history lengths. The final-six-layer intervention reproduced most of the mean effect of replacing all recurrent matrices. Adding convolution state changed neither the reported mean shift nor the count of positive pairs at the shown precision. Same-source copying produced zero shift. Together, these comparisons locate a reproducible contribution in the recurrent matrices while retaining the source's attention cache.

\begin{table*}[t]
\centering
\caption{Pre-query recurrent-state intervention}
\label{tab:intervention}
\small
\begin{tabular}{lrrr}
\toprule
Copied component & Mean shift (logits) & Profile-cluster 95\% interval & Positive pairs \\
\midrule
Same-source copy sham & \ResultInterventionShamMean{} & {[}\ResultInterventionShamLow{}, \ResultInterventionShamHigh{}{]} & \ResultInterventionShamPairs{} \\
All recurrent matrices & \ResultInterventionRecurrentMean{} & {[}\ResultInterventionRecurrentLow{}, \ResultInterventionRecurrentHigh{}{]} & \ResultInterventionRecurrentPairs{} \\
Final six recurrent layers & \ResultInterventionLateMean{} & {[}\ResultInterventionLateLow{}, \ResultInterventionLateHigh{}{]} & \ResultInterventionLatePairs{} \\
Recurrent plus convolution state & \ResultInterventionCombinedMean{} & {[}\ResultInterventionCombinedLow{}, \ResultInterventionCombinedHigh{}{]} & \ResultInterventionCombinedPairs{} \\
\bottomrule
\end{tabular}
\end{table*}

A complementary Granite intervention tested 96 ordered pairs at a snapshot after the public query. Replacing recurrent state shifted the donor-directed margin by \ResultGraniteInterventionState{} logits, with a profile-cluster interval of {[}\ResultGraniteInterventionStateLow{}, \ResultGraniteInterventionStateHigh{}{]}. Replacing recurrent and convolution state gave \ResultGraniteInterventionCombined{} {[}\ResultGraniteInterventionCombinedLow{}, \ResultGraniteInterventionCombinedHigh{}{]}. The snapshot differs from Qwen's stricter pre-query boundary, but the two models both showed a change in later decision margins when recurrent state was replaced. We use Qwen to establish the pre-query result and Granite to test the recurrent contribution in a second architecture. A single later public turn suffices: the victim's private context stays legible after the topic changes and still steers the model's next decision.

\subsection{Personal attributes across authors}\label{sec:5.3}

\begin{figure}[t]
\centering
\begin{tikzpicture}
\begin{axis}[
  paperplot, width=7.1cm, height=3.9cm,
  ymajorgrids=false, xmajorgrids, axis lines*=left,
  ymin=-0.5, ymax=2.5, ytick={0,1,2}, y dir=reverse, ytick style={draw=none},
  yticklabels from table={figdata/synthpai_selected.dat}{label},
  xmin=0, xmax=75, xtick={0,20,40,60}, xlabel={accuracy (\%)},
  legend style={at={(0.5,-0.35)}, anchor=north, legend columns=2, /tikz/every even column/.append style={column sep=6pt}},
  seg/.style={error bars/.cd, x dir=minus, x explicit, error mark=none,
    error bar style={line width=1.4pt, color=obsc!35}},
]
\addplot[only marks, mark=*, mark size=2pt, obsc, seg]
  table[x=latent,y=row,x error expr=\thisrow{latent}-\thisrow{base}] {figdata/synthpai_selected.dat};
\addplot[only marks, mark=o, mark size=2pt, black!65, line width=0.8pt]
  table[x=base,y=row] {figdata/synthpai_selected.dat};
\legend{state observation, fold majority}
\end{axis}
\end{tikzpicture}
\caption{Three personal-attribute tasks on the complete \ResultSynthAuthors{}-author SynthPAI corpus. Each comparison uses author-level out-of-fold predictions and the corresponding fold-training majority. The shown tasks pass a Holm correction across age and all eight evaluated categorical fields.}
\label{fig:4}
\end{figure}
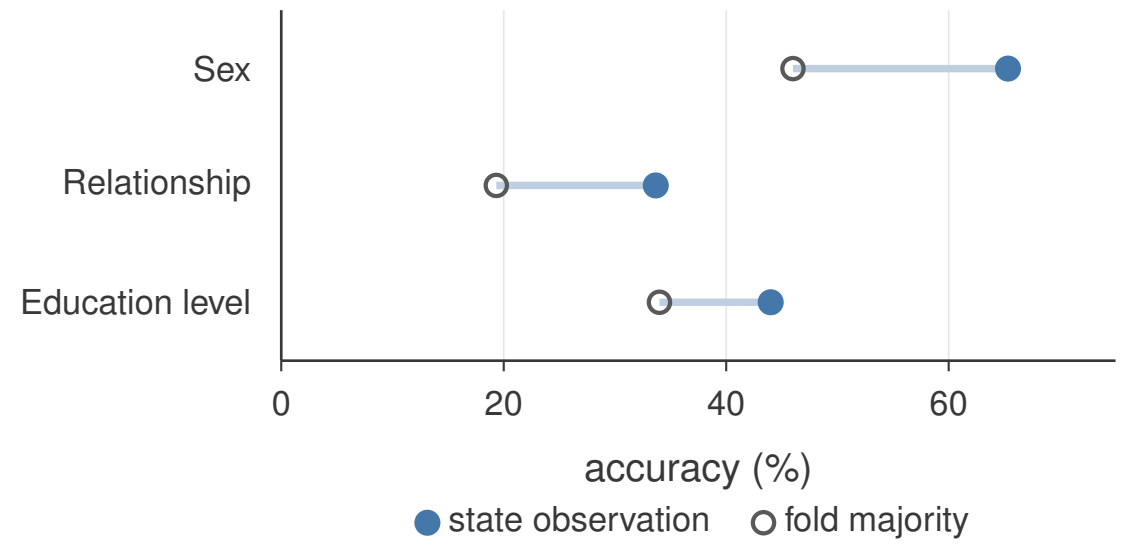

On all \ResultSynthAuthors{} SynthPAI authors, 8,192 sampled raw recurrent coordinates represented \ResultQwenFraction{}\% of Qwen state. Author-level out-of-fold age \ac{MAE} was \ResultSynthAgeMae{} years, compared with \ResultSynthAgeBaseline{} years for the fold-training median. Estimates fell within five years for \ResultSynthAgeWithinFive{}\% of authors, compared with \ResultSynthBaselineWithinFive{}\% for the baseline. Comparing per-author absolute errors gave a gain/loss count of \ResultSynthAgeGainLoss{}, with Holm-adjusted \(p=\ResultSynthAgeP{}\). Figure~\ref{fig:4} reports sex, relationship status, and education category. Each exceeded its fold-training majority and passed the Holm correction across age and all eight categorical fields. The labels are shared across authors, so these results measure generalization of attributes rather than recognition of a finite identity list.

The sampling procedure was label-blind: one seed-fixed coordinate set was used across the author population and task heads. The receiver received recurrent values, without author identifiers or comment text. Fold-training baselines accounted for the label distribution before each test fold, which matters for attributes with unequal class frequencies. Age evaluates prediction error across the released range; sex, relationship status, and education category ask whether shared labels generalize across authors with different text, all on the complete population. The full correction family was retained when reporting these attributes; selecting the displayed fields did not reduce the number of hypotheses used for correction. Each recovered field is a demographic attribute the adversary attaches to an author whose text it never reads.

\subsection{Power recovery from complete reads}\label{sec:5.4}

The captured model-derived read matched its floating-point reference with maximum absolute error \(\ResultOperatorError{}\). Quantized outputs correlated with the reference at \ResultQuantizationCorrelation{}, with normalized root mean square error \ResultQuantizationNrmse{}. The selected fixed-point operator was then evaluated through physical power traces on independent people. The victim-matched experiment recovered both private facts from complete-read traces (Table \ref{tab:cw}). Its actual-only receiver establishes task information without requiring paired traces at inference. The matched control retains coefficients derived from the same private history, allowing the paired comparison to test the additional contribution of the actual state input.

\begin{table*}[t]
\centering
\caption{Recovery of private facts from complete-read power traces}
\label{tab:cw}
\small
\begin{tabular}{lrrrr}
\toprule
Endpoint & Paired receiver & Actual-only receiver & Sham control & Paired vs.~sham, exact \(p\) \\
\midrule
\emph{Victim-matched arm, 64 unseen people} &  &  &  &  \\
Condition & \ResultCwNaturalPairedConditionCount{} & \ResultCwNaturalDirectConditionCount{} & \ResultCwNaturalShamConditionCount{} & \(\ResultCwNaturalConditionP{}\) \\
Vault word & \ResultCwNaturalPairedVaultCount{} & \ResultCwNaturalDirectVaultCount{} & \ResultCwNaturalShamVaultCount{} & \(\ResultCwNaturalVaultP{}\) \\
Both & \ResultCwNaturalPairedJointCount{} & \ResultCwNaturalDirectJointCount{} & \ResultCwNaturalShamJointCount{} & \(\ResultCwNaturalJointP{}\) \\
\emph{Fixed-coefficient arm, 64 unseen people} &  &  &  &  \\
Condition & \ResultCwFixedPairedConditionCount{} & \ResultCwFixedDirectConditionCount{} & \ResultCwFixedShamConditionCount{} & \(\ResultCwFixedConditionP{}\) \\
Vault word & \ResultCwFixedPairedVaultCount{} & \ResultCwFixedDirectVaultCount{} & \ResultCwFixedShamVaultCount{} & \(\ResultCwFixedVaultP{}\) \\
Both & \ResultCwFixedPairedJointCount{} & \ResultCwFixedDirectJointCount{} & \ResultCwFixedShamJointCount{} & \(\ResultCwFixedJointP{}\) \\
\bottomrule
\end{tabular}
\end{table*}

The fixed-coefficient experiment isolated recurrent state by keeping one model-derived coefficient vector for all people and providing public state at the sham's external input. Both receivers recovered private facts on a disjoint test population, every paired endpoint exceeding the public-input sham, which attributes the gain to confidential state under constant coefficients. Within-condition permutations gave \(p=\ResultCwConditionalP{}\) for vault and joint recovery under both receivers in each experiment. Across four separately averaged blocks, victim-matched condition accuracy ranged from \ResultCwNaturalConditionBlockRange{} and vault from \ResultCwNaturalVaultBlockRange{}; fixed-coefficient counts ranged from \ResultCwFixedConditionBlockRange{} and \ResultCwFixedVaultBlockRange{}. Reprogrammed bridges agreed with their profile-session predictions on all \ResultCwBridgePeople{} bridge people. Table \ref{tab:cwcost} reports the acquisition cost across actual and sham traces and all four test blocks.

\begin{table*}[t]
\centering
\caption{Complete-read acquisition cost, including actual and sham traces}
\label{tab:cwcost}
\small
\begin{tabular}{lrr}
\toprule
Session & Victim-matched (captures / s) & Fixed coefficient (captures / s) \\
\midrule
Profile & \ResultCostNaturalProfileCaptures{} / \ResultCostNaturalProfileSeconds{} & \ResultCostFixedProfileCaptures{} / \ResultCostFixedProfileSeconds{} \\
Bridge & \ResultCostNaturalBridgeCaptures{} / \ResultCostNaturalBridgeSeconds{} & \ResultCostFixedBridgeCaptures{} / \ResultCostFixedBridgeSeconds{} \\
Test & \ResultCostNaturalVictimCaptures{} / \ResultCostNaturalVictimSeconds{} & \ResultCostFixedVictimCaptures{} / \ResultCostFixedVictimSeconds{} \\
\bottomrule
\end{tabular}
\end{table*}

All profile, bridge, and terminal sessions completed without arithmetic-reference disagreements or capture timeouts, and the frozen receiver survived board reprogramming for known profiles. Neither the repeated blocks nor the bridge people entered the unseen-person count. The victim-matched arm measures coefficients and state produced together by a real invocation; the fixed-coefficient arm varies confidential state under one constant read vector with public bytes at the control input. Both yield private-fact inference, and the actual-only receivers show prediction can proceed from actual traces alone. One complete read of the state hands a physical adversary the user's secret and vault word, for people never used to build the receiver.

\subsection{Passive age inference on A100}\label{sec:5.5}

The routing experiment first tested whether its proposed intermediate affected execution. On a separate 32-target panel, all 128 all-anchor shams had \(Z=0\), while target batches spanned \ResultRoutingMin{}--\ResultRoutingMax{}. Singleton distances correlated with routing work at \(r=\ResultRoutingSingleton{}\). After subtracting anchor means, the correlation between routing work and the paired increase in CUDA execution time was \(r=\ResultRoutingResidual{}\) (target-block permutation \(p=\ResultRoutingResidualP{}\)). Instrumented routes and victim timing only evaluated the mechanism and were not observer inputs.

Independent validation connected routing work to duration and passive power transitions on unseen targets, meeting the prespecified criteria in Table \ref{tab:mechanism}. The sham-only comparison was consistent with no target association, separating state-dependent work from the anchor schedule. With validation targets disjoint from calibration, we ranked routing work and paired change-point shifts within each anchor and permuted assignments only within that anchor, testing target-dependent variation rather than a correlation among the four anchors. The mean paired change-point shift was \ResultMechanismShift{} seconds, and change-point timing estimated victim duration with \ac{MAE} \ResultMechanismMae{} seconds.

\begin{table*}[t]
\centering
\caption{Independent validation of the routing-work receiver}
\label{tab:mechanism}
\small
\begin{tabular}{lrrr}
\toprule
Measured relationship & Prespecified criterion & Spearman \(\rho\) & \(p\) \\
\midrule
Routing work vs.~paired victim duration & stratified \(\rho>.8\), \(p<.01\) & \(\ResultMechanismDuration{}\) & \(\ResultMechanismDurationP{}\) \\
Change point vs.~victim duration & \(\rho>.8\) & \(\ResultMechanismTracking{}\) & not prespecified \\
Routing work vs.~paired change-point shift & stratified \(\rho>.5\), \(p<.01\) & \(\ResultMechanismTransition{}\) & \(\ResultMechanismTransitionP{}\) \\
Routing work vs.~sham-only change point & \(p>.10\) & \(\ResultMechanismSham{}\) & \(\ResultMechanismShamP{}\) \\
\bottomrule
\end{tabular}
\end{table*}

The state-isolation terminal acquisition measured every test author against all four anchors, with \ResultAgePairs{} target/sham pairs over \ResultAgeRecords{} records including calibration (Figure~\ref{fig:5}). The receiver obtained four anchor distances from parent power; separately recorded GPU timing and route truth were used only to evaluate the measurement chain. Calibration correlations ranged from \ResultAgeCalibrationMin{} to \ResultAgeCalibrationMax{} across anchors. On terminal authors, physically estimated anchor distances correlated with their software values at \ResultAgeCoordinatesMin{}--\ResultAgeCoordinatesMax{}. Across all target/sham pairs, routing work correlated with the paired change-point shift at Spearman \(\rho=\ResultAgeWorkTransition{}\); change-point timing tracked victim duration at \(\rho=\ResultAgeTransitionDuration{}\).

\begin{figure*}[t]
\centering
\pgfplotstableread{figdata/routework_wave_cp.dat}\cptab
\pgfplotstablegetelem{0}{real}\of\cptab \let\cpreal\pgfplotsretval
\pgfplotstablegetelem{0}{sham}\of\cptab \let\cpsham\pgfplotsretval
\pgfplotstableread{figdata/routework_stats.dat}\sttab
\newcommand{\rwrho}[1]{\pgfplotstablegetelem{#1}{rho}\of\sttab Spearman $\rho$ = \pgfmathprintnumber[fixed, precision=3, assume math mode=true]{\pgfplotsretval}}
\subfloat[Parent power, one pair]{\begin{minipage}{.24\linewidth}\centering
\begin{tikzpicture}
\begin{axis}[paperplot, width=\linewidth, height=3.9cm,
  title style={font=\sffamily\footnotesize\bfseries, yshift=-1mm, align=center},
  anchorcol/.style={scatter, only marks, mark size=1.6pt,
    scatter src=explicit symbolic,
    scatter/classes={0={mark=*,draw=tolblue,fill=tolblue},
      1={mark=square*,draw=tolcyan,fill=tolcyan},
      2={mark=triangle*,draw=tolgreen,fill=tolgreen},
      3={mark=diamond*,draw=tolyellow!65!black,fill=tolyellow!65!black}}},
xlabel={time in record (s)}, ylabel={power (W)},
  xmin=20, xmax=32, xtick={20,24,28,32}, ymajorgrids=false,
  legend style={at={(0.03,0.03)}, anchor=south west}, legend cell align=left]

\addplot[secretc, line width=0.6pt] table[x=t, y=real] {figdata/routework_wave.dat};
\addplot[black!45, dashed, line width=0.6pt] table[x=t, y=sham] {figdata/routework_wave.dat};
\draw[secretc, dash pattern=on 2pt off 1.5pt] (axis cs:\cpreal,\pgfkeysvalueof{/pgfplots/ymin}) -- (axis cs:\cpreal,\pgfkeysvalueof{/pgfplots/ymax});
\draw[black!45, dash pattern=on 2pt off 1.5pt] (axis cs:\cpsham,\pgfkeysvalueof{/pgfplots/ymin}) -- (axis cs:\cpsham,\pgfkeysvalueof{/pgfplots/ymax});
\legend{target, sham}
\end{axis}
\end{tikzpicture}
\end{minipage}}\hfill
\subfloat[Victim duration]{\begin{minipage}{.24\linewidth}\centering
\begin{tikzpicture}
\begin{axis}[paperplot, width=\linewidth, height=3.9cm,
  title style={font=\sffamily\footnotesize\bfseries, yshift=-1mm, align=center},
  anchorcol/.style={scatter, only marks, mark size=1.6pt,
    scatter src=explicit symbolic,
    scatter/classes={0={mark=*,draw=tolblue,fill=tolblue},
      1={mark=square*,draw=tolcyan,fill=tolcyan},
      2={mark=triangle*,draw=tolgreen,fill=tolgreen},
      3={mark=diamond*,draw=tolyellow!65!black,fill=tolyellow!65!black}}},
xlabel={routing work $Z$}, ylabel={real $-$ sham (s)}, enlarge y limits={upper, value=0.3}]

\addplot[anchorcol] table[x=z, y=ddur, meta=anchor] {figdata/routework_pairs.dat};
\node[statnode] at (rel axis cs:0.02,0.98) {\rwrho{0}};
\end{axis}
\end{tikzpicture}
\end{minipage}}\hfill
\subfloat[Passive change point]{\begin{minipage}{.24\linewidth}\centering
\begin{tikzpicture}
\begin{axis}[paperplot, width=\linewidth, height=3.9cm,
  title style={font=\sffamily\footnotesize\bfseries, yshift=-1mm, align=center},
  anchorcol/.style={scatter, only marks, mark size=1.6pt,
    scatter src=explicit symbolic,
    scatter/classes={0={mark=*,draw=tolblue,fill=tolblue},
      1={mark=square*,draw=tolcyan,fill=tolcyan},
      2={mark=triangle*,draw=tolgreen,fill=tolgreen},
      3={mark=diamond*,draw=tolyellow!65!black,fill=tolyellow!65!black}}},
xlabel={routing work $Z$}, ylabel={real $-$ sham (s)}, enlarge y limits={upper, value=0.3}]

\addplot[anchorcol] table[x=z, y=dcp, meta=anchor] {figdata/routework_pairs.dat};
\node[statnode] at (rel axis cs:0.02,0.98) {\rwrho{1}};
\end{axis}
\end{tikzpicture}
\end{minipage}}\hfill
\subfloat[Change point vs. duration]{\begin{minipage}{.24\linewidth}\centering
\begin{tikzpicture}
\begin{axis}[paperplot, width=\linewidth, height=3.9cm,
  title style={font=\sffamily\footnotesize\bfseries, yshift=-1mm, align=center},
  anchorcol/.style={scatter, only marks, mark size=1.6pt,
    scatter src=explicit symbolic,
    scatter/classes={0={mark=*,draw=tolblue,fill=tolblue},
      1={mark=square*,draw=tolcyan,fill=tolcyan},
      2={mark=triangle*,draw=tolgreen,fill=tolgreen},
      3={mark=diamond*,draw=tolyellow!65!black,fill=tolyellow!65!black}}},
xlabel={victim duration (s)}, ylabel={change point (s)},
  enlarge y limits={upper, value=0.3}, legend style={at={(0.97,0.72)}, anchor=north east}, legend cell align=left]

\addplot[only marks, mark=square*, mark size=1.1pt, black!40, restrict expr to domain={\thisrow{real}}{0:0}] table[x=dur, y=cp] {figdata/routework_records.dat};
\addplot[only marks, mark=*, mark size=1.1pt, secretc, restrict expr to domain={\thisrow{real}}{1:1}] table[x=dur, y=cp] {figdata/routework_records.dat};
\node[statnode] at (rel axis cs:0.02,0.98) {\rwrho{2}};
\legend{sham, target}
\end{axis}
\end{tikzpicture}
\end{minipage}}
\par\smallskip{\footnotesize Anchors: $\bullet$ 1\quad $\blacksquare$ 2\quad $\blacktriangle$ 3\quad $\blacklozenge$ 4.}
\caption{Passive routing-work mechanism in the state-isolation run, all \ResultAgePairs{} real/sham pairs and \ResultAgeRecords{} records. (a) Parent power for the pair with the median \(Z\) and its sham, with the frozen receiver's change points dashed. (b) Routing work \(Z\) versus the paired increase in victim duration. (c) \(Z\) versus the paired change-point shift. (d) Change point versus victim duration for every record. Point shapes in (b) and (c) distinguish the public anchors; \(\rho\) is the recorded Spearman correlation.}
\label{fig:5}
\end{figure*}

On the \ResultAgeAuthors{} held-out authors, the age head on clean software distances had age \ac{MAE} \ResultAgeClean{} years. Recovered anchor distances gave \ResultAgePhysical{} years, compared with the frozen training-median baseline of \ResultAgeBaseline{} years (one-sided paired sign-flip \(p=\ResultAgeP{}\)). The physical and software predictions correlated at \(r=\ResultAgeCorrelation{}\). Applying the same calibration procedure to sham-only measurements gave \ac{MAE} \ResultAgeSham{} years. The supplementary bootstrap gave a 95\% interval of {[}\ResultAgePhysicalCiLow{}, \ResultAgePhysicalCiHigh{}{]} years for physical \ac{MAE}. The paired improvement over the median baseline was \ResultAgeGain{} years, with interval {[}\ResultAgeGainCiLow{}, \ResultAgeGainCiHigh{}{]}. The matched comparison attributes the prediction gain to the state-dependent routing signal.

The state-isolation acquisition used \ResultAgeTotalHours{} hours of scheduled slots, or \ResultAgeSecondsPerAuthor{} s per terminal author. Its passive age inference requires repeated public execution, anchors, same-session calibration, and exposed parent power.

\begin{figure*}[t]
\centering
\pgfplotstableread{figdata/age_stats.dat}\agetab
\pgfplotstablegetelem{0}{phys}\of\agetab \let\agephys\pgfplotsretval
\pgfplotstablegetelem{0}{clean}\of\agetab \let\ageclean\pgfplotsretval
\pgfplotstablegetelem{0}{base}\of\agetab \let\agebase\pgfplotsretval
\pgfplotstablegetelem{0}{r}\of\agetab \let\ager\pgfplotsretval
\pgfmathprintnumberto[fixed, fixed zerofill, precision=2, assume math mode=true]{\agephys}{\agephysS}
\pgfmathprintnumberto[fixed, fixed zerofill, precision=2, assume math mode=true]{\ageclean}{\agecleanS}
\subfloat[Four recovered anchor distances]{\begin{minipage}{.32\linewidth}\centering
\begin{tikzpicture}
\begin{axis}[paperplot, width=\linewidth, height=4.4cm,
  title style={font=\sffamily\footnotesize\bfseries, yshift=-1mm},
  anchorcol/.style={scatter, only marks, mark size=1.6pt,
    scatter src=explicit symbolic,
    scatter/classes={0={mark=*,draw=tolblue,fill=tolblue},
      1={mark=square*,draw=tolcyan,fill=tolcyan},
      2={mark=triangle*,draw=tolgreen,fill=tolgreen},
      3={mark=diamond*,draw=tolyellow!65!black,fill=tolyellow!65!black}}},
  diag/.style={negc, dash pattern=on 2pt off 1.5pt, line width=0.6pt, domain=200:460, samples=2},
xlabel={true anchor distance}, ylabel={recovered from power},
  xmin=200, xmax=460, ymin=200, ymax=460, xtick={200,300,400}, ytick={200,300,400}]

\addplot[diag, forget plot] {x};
\addplot[anchorcol] table[x=true, y=est, meta=anchor] {figdata/age_coords.dat};
\end{axis}
\end{tikzpicture}
\end{minipage}}\hfill
\subfloat[Age prediction]{\begin{minipage}{.32\linewidth}\centering
\begin{tikzpicture}
\begin{axis}[paperplot, width=\linewidth, height=4.4cm,
  title style={font=\sffamily\footnotesize\bfseries, yshift=-1mm},
  anchorcol/.style={scatter, only marks, mark size=1.6pt,
    scatter src=explicit symbolic,
    scatter/classes={0={mark=*,draw=tolblue,fill=tolblue},
      1={mark=square*,draw=tolcyan,fill=tolcyan},
      2={mark=triangle*,draw=tolgreen,fill=tolgreen},
      3={mark=diamond*,draw=tolyellow!65!black,fill=tolyellow!65!black}}},
  diag/.style={negc, dash pattern=on 2pt off 1.5pt, line width=0.6pt, domain=200:460, samples=2},
xlabel={clean software prediction (years)}, ylabel={physical prediction (years)},
  xmin=20, xmax=60, ymin=20, ymax=60]

\addplot[negc, dash pattern=on 2pt off 1.5pt, line width=0.6pt, domain=20:60, samples=2, forget plot] {x};
\addplot[only marks, mark=*, mark size=1.5pt, obsc, fill opacity=0.85, draw=white, line width=0.3pt] table[x=clean, y=phys] {figdata/age_pred.dat};
\node[statnode] at (rel axis cs:0.02,0.98) {$r$ = \pgfmathprintnumber[fixed, precision=3, assume math mode=true]{\ager}};
\end{axis}
\end{tikzpicture}
\end{minipage}}\hfill
\subfloat[Absolute age error]{\begin{minipage}{.32\linewidth}\centering
\begin{tikzpicture}
\begin{axis}[paperplot, width=\linewidth, height=4.4cm,
  title style={font=\sffamily\footnotesize\bfseries, yshift=-1mm},
  anchorcol/.style={scatter, only marks, mark size=1.6pt,
    scatter src=explicit symbolic,
    scatter/classes={0={mark=*,draw=tolblue,fill=tolblue},
      1={mark=square*,draw=tolcyan,fill=tolcyan},
      2={mark=triangle*,draw=tolgreen,fill=tolgreen},
      3={mark=diamond*,draw=tolyellow!65!black,fill=tolyellow!65!black}}},
  diag/.style={negc, dash pattern=on 2pt off 1.5pt, line width=0.6pt, domain=200:460, samples=2},
xlabel={absolute error (years)}, ylabel={fraction of authors},
  xmin=0, xmax=42, ymin=0, ymax=1.02, ytick={0,0.25,0.5,0.75,1},
  legend style={at={(0.97,0.03)}, anchor=south east}, legend cell align=left]

\addplot[obsc, const plot, line width=1pt] table[x=phys, y=q] {figdata/age_err.dat};
\addplot[black!50, dashed, const plot, line width=0.8pt] table[x=clean, y=q] {figdata/age_err.dat};
\addplot[negc, dash pattern=on 2pt off 1.5pt, line width=0.6pt] coordinates {(\agebase,0) (\agebase,1.02)};
\legend{{physical (MAE \agephysS)}, {clean (MAE \agecleanS)}, {median baseline MAE}}
\end{axis}
\end{tikzpicture}
\end{minipage}}
\par\smallskip{\footnotesize Anchors: $\bullet$ 1\quad $\blacksquare$ 2\quad $\blacktriangle$ 3\quad $\blacklozenge$ 4.}
\caption{State-isolated passive age inference on all \ResultAgeAuthors{} fold-4 authors. (a) Anchor distance for each of the four public anchors recovered from parent-power change points versus its true value; the dashed line is equality. (b) Age estimates of the frozen head from recovered anchor distances versus from clean software distances. (c) Empirical distribution of absolute age error for both, with the frozen training-median baseline \acs{MAE} dashed; the sham-only receiver, not drawn, has \acs{MAE} \ResultAgeSham{} years.}
\label{fig:6}
\end{figure*}
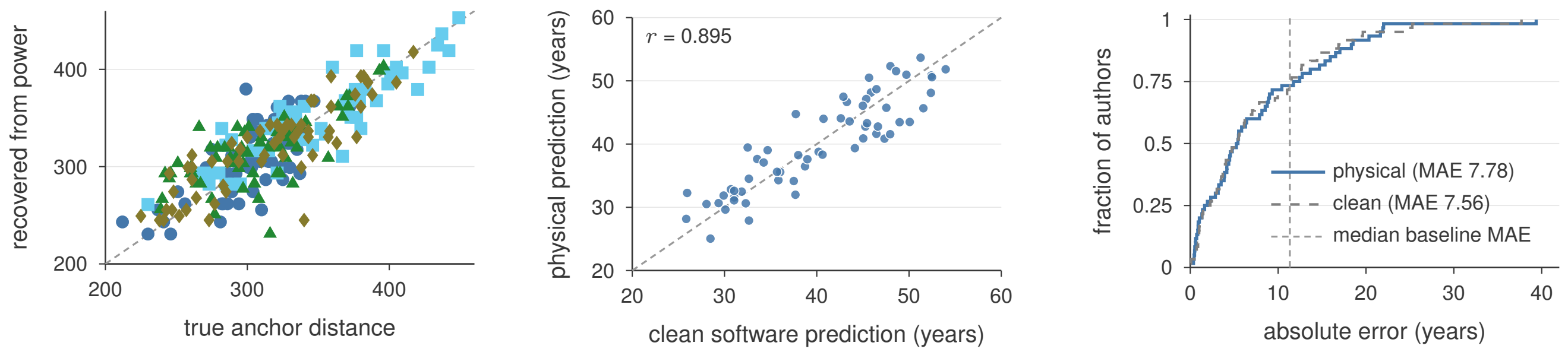

The natural-session endpoint tests whether the public request and physical observation can belong to one actor: the host client submitted every continuation with opaque handles and sampled parent power, while the service retained complete private caches and returned no response text. A post-acquisition instrumented replay reproduced every victim checksum and counted exactly 1,920 active expert groups in all \ResultNaturalAgePairs{} shams. Actual mixed-batch excess work correlated with paired victim-duration shift at \(\rho=\ResultNaturalBatchWorkDurationRho{}\) and parent-power change-point shift at \(\rho=\ResultNaturalBatchWorkPowerRho{}\); within-anchor work-to-power correlations ranged from \ResultNaturalBatchAnchorRhoMin{} to \ResultNaturalBatchAnchorRhoMax{}. The passive change point tracked victim duration at \(\rho=\ResultNaturalAgePowerDurationRho{}\). On all \ResultNaturalAgeAuthors{} terminal authors, the frozen receiver reached \ResultNaturalAgePhysical{} years age \ac{MAE}, compared with \ResultNaturalAgeBaseline{} years for the training-median baseline (one-sided paired sign-flip \(p=\ResultNaturalAgeP{}\)). Clean natural-cache distances gave \ResultNaturalAgeClean{} years, and the sham-only receiver gave \ResultNaturalAgeSham{} years. This run links one client's request, parent-power observation, and attribute prediction under an append-only session API, reusing the state-isolation terminal authors with changed cache construction and observer access and no terminal label in the client or calibration; the state-isolation run supplies the recurrent-specific attribution. In the deployment this models, one unprivileged account on the shared GPU recovers a tenant's personal attribute from parent power alone.

\subsection{Implications for context confidentiality}\label{sec:5.6}

Our results establish a flow from retained private context to observations of its computation: the content stayed inferable after public dialogue and kept shifting decision margins after the conversation moved on. Applications that keep private profiles or instructions in recurrent state across requests carry this flow, which changes three deployment decisions.

Serving systems that cache, offload, migrate, or share recurrent state should give it the access control and isolation of the prompt that produced it and clear or recompute it at session boundaries, both within the serving software. GPU cloud operators that partition devices among tenants should treat whole-device parent power as tenant data: in the natural-session service an unprivileged account inferred age from it, and a non-root \ac{MIG} container read the same signal, neither needing a model change. Accelerator designers should note that fixed control flow removed schedule dependence while data-dependent switching still disclosed private facts, so a constant-schedule datapath does not protect confidential state.

\section{Conclusion}\label{sec:6}

We showed that the recurrent state of latent-state LLMs carries confidential prompt content into later computation. Partial, noisy observations recovered private facts, response rules, and personal attributes of unseen people, and this information persisted through public dialogue and reached physical observers through power. As these designs spread to bound serving memory, the confidentiality of a personalized session comes to depend on the recurrent state, its lifetime, and its power, which serving systems, GPU operators, and accelerator designers each partly control.

\label{main:lasttext} \clearpage \appendices

\section{Acquisition and replay details}\label{app:A}

The CW donor-selection procedure ranks candidates using the smaller of condition and vault validation accuracy, then their sum, then donor index. Only 384 core profiles participate. The selected donor is \ResultCwDonor{}. The fixed-coefficient control retains the same complete 128-term reads and supplies eight fixed public bytes at its external input. This removes real-state loading from the control path. Each physical record is checked against an independent integer arithmetic reference. Capture counts include both actual and sham arms. Profile, bridge, and test processes reprogram the board separately. Control assignments and randomized record order are fixed in the acquisition manifests, which bind the operands, bitstreams, waveforms, and receiver outputs.

The A100 acquisition snapshot dated \ResultAgeDate{} (UTC) also binds the victim code, model configuration, and weight-index file. Replay of the reported analyses starts from the retained operands, routes, and predictions; new extraction pins a model revision. A100 manifests bind the frozen schedule, routing truth, victim log, and observer samples by hash. The terminal run retained all \ResultAgeRecords{} victim records and \ResultAgeSamples{} power samples. The recorded rule carried forward \ResultAgeInvalidSamples{} invalid power reads. A 12 GiB idle allocation reserved during a probe prevented an unrelated same-account workload from entering the victim instance. The receiver used parent power; the victim log was joined only during analysis.

The natural-session manifest binds the service and client code, opaque request schedule, clean route set, victim log, and parent-power samples. The run retained \ResultNaturalAgeRecords{} victim records and \ResultNaturalAgeSamples{} power samples; \ResultNaturalAgeInvalidSamples{} invalid telemetry readings followed the frozen forward-fill rule. The service preserved all four components of each naturally prefetched cache, while the host client had session handles with append-only access and parent-power access. Its verifier checks the service, client, and observer timing windows before scoring the frozen receiver.

Replay entry points cover complete-read, fixed-coefficient, and both A100 age endpoints. They recompute predictions and paired statistics from retained records, validate the acquisition manifests, and reproduce the figure data. Register-transfer-level replay checks the quantized read arithmetic and public control schedule. Experimental development histories remain in the laboratory record; they are not pooled into the terminal test populations.

\bibliographystyle{IEEEtran-preserve}
\bibliography{refs}

\end{document}